\font\twelverm=cmr10 scaled 1200    \font\twelvei=cmmi10 scaled 1200
\font\twelvesy=cmsy10 scaled 1200   \font\twelveex=cmex10 scaled 1200
\font\twelvebf=cmbx10 scaled 1200   \font\twelvesl=cmsl10 scaled 1200
\font\twelvett=cmtt10 scaled 1200   \font\twelveit=cmti10 scaled 1200
\skewchar\twelvei='177   \skewchar\twelvesy='60
\def\twelvepoint{\normalbaselineskip=12.4pt
  \abovedisplayskip 12.4pt plus 3pt minus 9pt
  \belowdisplayskip 12.4pt plus 3pt minus 9pt
  \abovedisplayshortskip 0pt plus 3pt
  \belowdisplayshortskip 7.2pt plus 3pt minus 4pt
  \smallskipamount=3.6pt plus1.2pt minus1.2pt
  \medskipamount=7.2pt plus2.4pt minus2.4pt
  \bigskipamount=14.4pt plus4.8pt minus4.8pt
  \def\rm{\fam0\twelverm}          \def\it{\fam\itfam\twelveit}%
  \def\sl{\fam\slfam\twelvesl}     \def\bf{\fam\bffam\twelvebf}%
  \def\mit{\fam 1}                 \def\cal{\fam 2}%
  \def\tt{\twelvett}
  \textfont0=\twelverm   \scriptfont0=\tenrm   \scriptscriptfont0=\sevenrm
  \textfont1=\twelvei    \scriptfont1=\teni    \scriptscriptfont1=\seveni
  \textfont2=\twelvesy   \scriptfont2=\tensy   \scriptscriptfont2=\sevensy
  \textfont3=\twelveex   \scriptfont3=\twelveex  \scriptscriptfont3=\twelveex
  \textfont\itfam=\twelveit
  \textfont\slfam=\twelvesl
  \textfont\bffam=\twelvebf \scriptfont\bffam=\tenbf
  \scriptscriptfont\bffam=\sevenbf
  \normalbaselines\rm}

\def\beginlinemode{\endmode
  \begingroup\parskip=0pt \obeylines\def\\{\par}\def\endmode{\par\endgroup}}
\def\beginparmode{\endmode
  \begingroup \def\endmode{\par\endgroup}}
\let\endmode=\par
{\obeylines\gdef\
{}}
\def\singlespace{\baselineskip=\normalbaselineskip}

\def\oneandahalfspace{\baselineskip=\normalbaselineskip
  \multiply\baselineskip by 3 \divide\baselineskip by 2}
\def\doublespace{\baselineskip=\normalbaselineskip \multiply\baselineskip by 2}
\newcount\firstpageno
\firstpageno=2
\footline={\ifnum\pageno<\firstpageno{\hfil}%
\else{\hfil\twelverm\folio\hfil}\fi}
\let\rawfootnote=\footnote              
\def\footnote#1#2{{\rm\singlespace\parindent=0pt\rawfootnote{#1}{#2}}}
\def\raggedcenter{\leftskip=2em plus 12em \rightskip=\leftskip
  \parindent=0pt \parfillskip=0pt \spaceskip=.3333em \xspaceskip=.5em
  \pretolerance=9999 \tolerance=9999
  \hyphenpenalty=9999 \exhyphenpenalty=9999 }
\parskip=\medskipamount
\twelvepoint            
\overfullrule=0pt       
\def\preprintno#1{
 \rightline{\rm #1}}    
\def\author                     
  {\vskip 3pt plus 0.2fill \beginlinemode
   \singlespace \raggedcenter \twelvesc}
\def\affil                      
  {\vskip 3pt plus 0.1fill \beginlinemode
   \oneandahalfspace \raggedcenter \sl}
\def\abstract                   
  {\vskip 3pt plus 0.3fill \beginparmode
   \doublespace \narrower \noindent ABSTRACT: }
\def\endtitlepage               
  {\endpage                     
   \body}
\def\body                       
  {\beginparmode}               

\def\subhead#1{                 
  \vskip 0.25truein             
  {\raggedcenter #1 \par}
   \nobreak\vskip 0.1truein\nobreak}
\def\refto#1{$|{#1}$}           
\def\references                
  {\subhead{References}        
   \beginparmode
   \frenchspacing \parindent=0pt \leftskip=1truecm
   \parskip=8pt plus 3pt \everypar{\hangindent=\parindent}}
\gdef\refis#1{\indent\hbox to 0pt{\hss#1.~}}    
\gdef\journal#1, #2, #3, 1#4#5#6{               
    {\sl #1~}{\bf #2}, #3, (1#4#5#6)}           
\def\refstylenp{                
  \gdef\refto##1{ [##1]}                                
  \gdef\refis##1{\indent\hbox to 0pt{\hss##1)~}}        
  \gdef\journal##1, ##2, ##3, ##4 {                     
     {\sl ##1~}{\bf ##2~}(##3) ##4 }}
\def\refstyleprnp{              
  \gdef\refto##1{ [##1]}                                
  \gdef\refis##1{\indent\hbox to 0pt{\hss##1)~}}        
  \gdef\journal##1, ##2, ##3, 1##4##5##6{               
    {\sl ##1~}{\bf ##2~}(1##4##5##6) ##3}}
\def\pr{\journal Phys. Rev., }
\def\prd{\journal Phys. Rev. D, }
\def\prl{\journal Phys. Rev. Lett., }
\def\prpts{\journal Phys. Rep., }
\def\np{\journal Nucl. Phys., }
\def\pl{\journal Phys. Lett., }

\def\endreferences{\body}
\def\endpage                    
  {\vfill\eject}
\def\endpaper                   
  {\endmode\vfill\supereject}
\def\endit
  {\endpaper\end}
\def\ref#1{Ref. #1}                     
\def\Ref#1{Ref. #1}                     

\def\m@th{\mathsurround=0pt }
\font\twelvesc=cmcsc10 scaled 1200
\def\cite#1{{#1}}
\def\(#1){(\call{#1})}
\def\call#1{{#1}}
\def\taghead#1{}
\def\leaderfill{\leaders\hbox to 1em{\hss.\hss}\hfill}
\def\twiddle{\lower.9ex\rlap{$\kern-.1em\scriptstyle\sim$}}
\def\bigtwiddle{\lower1.ex\rlap{$\sim$}}
\def\gtwid{\mathrel{\raise.3ex\hbox{$>$\kern-.75em\lower1ex\hbox{$\sim$}}}}
\def\ltwid{\mathrel{\raise.3ex\hbox{$<$\kern-.75em\lower1ex\hbox{$\sim$}}}}
\def\square{\kern1pt\vbox{\hrule height 1.2pt\hbox{\vrule width 1.2pt\hskip 3pt
   \vbox{\vskip 6pt}\hskip 3pt\vrule width 0.6pt}\hrule height 0.6pt}\kern1pt}
\catcode`@=11
\newcount\tagnumber\tagnumber=0

\immediate\newwrite\eqnfile
\newif\if@qnfile\@qnfilefalse
\def\write@qn#1{}
\def\writenew@qn#1{}
\def\w@rnwrite#1{\write@qn{#1}\message{#1}}
\def\@rrwrite#1{\write@qn{#1}\errmessage{#1}}

\def\taghead#1{\gdef\t@ghead{#1}\global\tagnumber=0}
\def\t@ghead{}

\expandafter\def\csname @qnnum-3\endcsname
  {{\t@ghead\advance\tagnumber by -3\relax\number\tagnumber}}
\expandafter\def\csname @qnnum-2\endcsname
  {{\t@ghead\advance\tagnumber by -2\relax\number\tagnumber}}
\expandafter\def\csname @qnnum-1\endcsname
  {{\t@ghead\advance\tagnumber by -1\relax\number\tagnumber}}
\expandafter\def\csname @qnnum0\endcsname
  {\t@ghead\number\tagnumber}
\expandafter\def\csname @qnnum+1\endcsname
  {{\t@ghead\advance\tagnumber by 1\relax\number\tagnumber}}
\expandafter\def\csname @qnnum+2\endcsname
  {{\t@ghead\advance\tagnumber by 2\relax\number\tagnumber}}
\expandafter\def\csname @qnnum+3\endcsname
  {{\t@ghead\advance\tagnumber by 3\relax\number\tagnumber}}

\def\equationfile{%
  \@qnfiletrue\immediate\openout\eqnfile=\jobname.eqn%
  \def\write@qn##1{\if@qnfile\immediate\write\eqnfile{##1}\fi}
  \def\writenew@qn##1{\if@qnfile\immediate\write\eqnfile
    {\noexpand\tag{##1} = (\t@ghead\number\tagnumber)}\fi}
}

\def\callall#1{\xdef#1##1{#1{\noexpand\call{##1}}}}
\def\call#1{\each@rg\callr@nge{#1}}

\def\each@rg#1#2{{\let\thecsname=#1\expandafter\first@rg#2,\end,}}
\def\first@rg#1,{\thecsname{#1}\apply@rg}
\def\apply@rg#1,{\ifx\end#1\let\next=\relax%
\else,\thecsname{#1}\let\next=\apply@rg\fi\next}

\def\callr@nge#1{\calldor@nge#1-\end-}
\def\callr@ngeat#1\end-{#1}
\def\calldor@nge#1-#2-{\ifx\end#2\@qneatspace#1 %
  \else\calll@@p{#1}{#2}\callr@ngeat\fi}
\def\calll@@p#1#2{\ifnum#1>#2{\@rrwrite{Equation range #1-#2\space is bad.}
\errhelp{If you call a series of equations by the notation M-N, then M and
N must be integers, and N must be greater than or equal to M.}}\else%
{\count0=#1\count1=#2\advance\count1 by1\relax\expandafter\@qncall\the\count0,%
  \loop\advance\count0 by1\relax%
    \ifnum\count0<\count1,\expandafter\@qncall\the\count0,%
  \repeat}\fi}

\def\@qneatspace#1#2 {\@qncall#1#2,}
\def\@qncall#1,{\ifunc@lled{#1}{\def\next{#1}\ifx\next\empty\else
  \w@rnwrite{Equation number \noexpand\(>>#1<<) has not been defined yet.}
  >>#1<<\fi}\else\csname @qnnum#1\endcsname\fi}

\let\eqnono=\eqno
\def\eqno(#1){\tag#1}
\def\tag#1$${\eqnono(\displayt@g#1 )$$}

\def\aligntag#1\endaligntag
  $${\gdef\tag##1\\{&(##1 )\cr}\eqalignno{#1\\}$$
  \gdef\tag##1$${\eqnono(\displayt@g##1 )$$}}

\def\eqalignno#1{\displ@y \tabskip\centering
  \halign to\displaywidth{\hfil$\displaystyle{##}$\tabskip\z@skip
    &$\displaystyle{{}##}$\hfil\tabskip\centering
    &\llap{$\displayt@gpar##$}\tabskip\z@skip\crcr
    #1\crcr}}

\def\displayt@gpar(#1){(\displayt@g#1 )}

\def\displayt@g#1 {\rm\ifunc@lled{#1}\global\advance\tagnumber by1
        {\def\next{#1}\ifx\next\empty\else\expandafter
        \xdef\csname @qnnum#1\endcsname{\t@ghead\number\tagnumber}\fi}%
  \writenew@qn{#1}\t@ghead\number\tagnumber\else
        {\edef\next{\t@ghead\number\tagnumber}%
        \expandafter\ifx\csname @qnnum#1\endcsname\next\else
        \w@rnwrite{Equation \noexpand\tag{#1} is a duplicate number.}\fi}%
  \csname @qnnum#1\endcsname\fi}

\def\ifunc@lled#1{\expandafter\ifx\csname @qnnum#1\endcsname\relax}

\let\@qnend=\end\gdef\end{\if@qnfile
\immediate\write16{Equation numbers written on []\jobname.EQN.}\fi\@qnend}

\catcode`@=12
\catcode`@=11
\newcount\r@fcount \r@fcount=0
\def\refreset{\global\r@fcount=0}
\newcount\r@fcurr
\immediate\newwrite\reffile
\newif\ifr@ffile\r@ffilefalse
\def\w@rnwrite#1{\ifr@ffile\immediate\write\reffile{#1}\fi\message{#1}}

\def\writer@f#1>>{}
\def\referencefile{
  \r@ffiletrue\immediate\openout\reffile=\jobname.ref%
  \def\writer@f##1>>{\ifr@ffile\immediate\write\reffile%
    {\noexpand\refis{##1} = \csname r@fnum##1\endcsname = %
     \expandafter\expandafter\expandafter\strip@t\expandafter%
     \meaning\csname r@ftext\csname r@fnum##1\endcsname\endcsname}\fi}%
  \def\strip@t##1>>{}}

\def\citeall#1{\xdef#1##1{#1{\noexpand\cite{##1}}}}
\def\cite#1{\each@rg\citer@nge{#1}}

\def\each@rg#1#2{{\let\thecsname=#1\expandafter\first@rg#2,\end,}}
\def\first@rg#1,{\thecsname{#1}\apply@rg}	
\def\apply@rg#1,{\ifx\end#1\let\next=\relax
\else,\thecsname{#1}\let\next=\apply@rg\fi\next}

\def\citer@nge#1{\citedor@nge#1-\end-}	
\def\citer@ngeat#1\end-{#1}
\def\citedor@nge#1-#2-{\ifx\end#2\r@featspace#1 
  \else\citel@@p{#1}{#2}\citer@ngeat\fi}	
\def\citel@@p#1#2{\ifnum#1>#2{\errmessage{Reference range #1-#2\space is bad.}%
    \errhelp{If you cite a series of references by the notation M-N, then M and
    N must be integers, and N must be greater than or equal to M.}}\else%
{\count0=#1\count1=#2\advance\count1 by1\relax\expandafter\r@fcite\the\count0,%
  \loop\advance\count0 by1\relax
    \ifnum\count0<\count1,\expandafter\r@fcite\the\count0,%
  \repeat}\fi}

\def\r@featspace#1#2 {\r@fcite#1#2,} 
\def\r@fcite#1,{\ifuncit@d{#1}
    \newr@f{#1}%
    \expandafter\gdef\csname r@ftext\number\r@fcount\endcsname%
                     {\message{Reference #1 to be supplied.}%
                      \writer@f#1>>#1 to be supplied.\par}%
 \fi%
 \csname r@fnum#1\endcsname}
\def\ifuncit@d#1{\expandafter\ifx\csname r@fnum#1\endcsname\relax}%
\def\newr@f#1{\global\advance\r@fcount by1%
    \expandafter\xdef\csname r@fnum#1\endcsname{\number\r@fcount}}

\let\r@fis=\refis			
\def\refis#1#2#3\par{\ifuncit@d{#1}
   \newr@f{#1}%
  \w@rnwrite{Reference #1=\number\r@fcount\space is not cited up to now.}\fi%
  \expandafter\gdef\csname r@ftext\csname r@fnum#1\endcsname\endcsname%
  {\writer@f#1>>#2#3\par}}

\def\ignoreuncited{
   \def\refis##1##2##3\par{\ifuncit@d{##1}%
    \else\expandafter\gdef\csname r@ftext\csname r@fnum##1\endcsname\endcsname%
     {\writer@f##1>>##2##3\par}\fi}}

\def\r@ferr{\endreferences\errmessage{I was expecting to see
\noexpand\endreferences before now;  I have inserted it here.}}
\let\r@ferences=\references
\def\references{\r@ferences\def\endmode{\r@ferr\par\endgroup}}

\let\endr@ferences=\endreferences
\def\endreferences{\r@fcurr=0
  {\loop\ifnum\r@fcurr<\r@fcount
   \advance\r@fcurr by 1\relax\expandafter\r@fis\expandafter{\number\r@fcurr}%
    \csname r@ftext\number\r@fcurr\endcsname%
  \repeat}\gdef\r@ferr{}\global\r@fcount=0\endr@ferences}

\let\r@fend=\endpaper\gdef\endpaper{\ifr@ffile
\immediate\write16{Cross References written on []\jobname.REF.}\fi\r@fend}

\catcode`@=12

\citeall\refto		
\citeall\ref		%
\citeall\Ref		%

\referencefile
\def\sss{{\cal S}^\prime}
\def\msbar{\overline{\rm MS}}
\def\drbar{\overline{\rm DR}}
\def\b{b}
\def\byuk{{\bf Y}}
\def\bsh{{\bf h}}
\def\mhu{{m^2_{H_u}}}
\def\mhd{{m^2_{H_d}}}
\def\bmq{{\bf m}^2_Q}
\def\bml{{\bf m}^2_L}
\def\bmu{{\bf m}^2_u}
\def\bmd{{\bf m}^2_d}
\def\bme{{\bf m}^2_e}
\def\mij{{(m^2)}_i^j}
\def\ssshu{{\hat H}_u}
\def\ssshd{{\hat H}_d}
\def\sssq{{\hat Q}}
\def\sssl{{\hat L}}
\def\sssd{{\hat d}}
\def\sssu{{\hat u}}
\def\ssse{{\hat e}}
\def\trym{{\cal S}}

\def\ss{\scriptscriptstyle}
\def\frac#1/#2{#1 / #2}
\def\neuphys{Department of Physics\\Northeastern University\\Boston MA 02115}
\def\oneandfourfifthsspace{\baselineskip=\normalbaselineskip
  \multiply\baselineskip by 9 \divide\baselineskip by 5}

\font\titlefont=cmr10 scaled\magstep3
\def\bigtitle                      
  {\null\vskip 3pt plus 0.2fill
   \beginlinemode \doublespace \raggedcenter \titlefont}

\oneandfourfifthsspace
\preprintno{NUB-3081-93TH}
\preprintno{hep-ph 9311340}
\preprintno{November 1993}
\preprintno{(revised June 1994)}
\bigtitle{Two-Loop Renormalization Group Equations
          for Soft Supersymmetry-Breaking Couplings}
\author Stephen P. Martin and Michael T. Vaughn
\affil\neuphys
\body

\abstract
We compute the two-loop renormalization group equations for all soft
supersymmetry-breaking couplings in a general softly broken $N=1$
supersymmetric model. We also specialize these results to the
Minimal Supersymmetric Standard Model.

\endtitlepage

\subhead{I. Introduction}
\taghead{1.}

In the Standard Model, the mass of the Higgs scalar boson is subject to
quadratically divergent radiative corrections which ought to be of order
$\delta m_H^2 \sim \Lambda^2$, where $\Lambda$ is an ultraviolet cutoff
scale. This presents a naturalness problem if $\Lambda^2 \gg m_H^2$, since
fine tuning is then necessary to explain why the Higgs mass is near the
electroweak scale. Low-energy supersymmetry[\cite{reviews}]
evades this naturalness problem because the quadratic divergences cancel.
To accomplish this, the Minimal Supersymmetric Standard Model (MSSM)
introduces a ``sparticle" partner for each standard model particle.
If supersymmetry were exact each sparticle would be degenerate in mass with
its standard model partner, which is certainly not the case experimentally,
so supersymmetry must be broken. Fortunately, the cancellation of quadratic
divergences still works if supersymmetry is broken softly[\cite{softly}]
by terms of dimension 2 and 3 in the Lagrangian. The masses of the
sparticles are then determined by the soft supersymmetry-breaking terms.

Because low-energy supersymmetry is a perturbative solution to the naturalness
problem, one can attempt to relate observed phenomena at low energies to
physics at very high energy scales. For example, it is remarkable that in
the MSSM, the three gauge couplings appear to unify[\cite{unification}]
at a scale $10^{15}$-$10^{16}$ GeV, hinting at a Grand Unified Theory (GUT)
or some other organizing principle such as string theory.
With the eventual discovery
of the sparticles and determination of their masses, we should gain
information about the soft supersymmetry-breaking parameters. Already
we know that these parameters are not at all arbitrary, because otherwise
large flavor-changing neutral currents would arise in the low-energy
physics due to the effects of loops containing squarks, and the arbitrary
complex phases of the soft supersymmetry-breaking terms would be
expected to give rise to a CP-violating electric dipole moment for the
neutron in violation of experimental bounds. Thus there is strong
circumstantial evidence in favor of some organizing principle governing
the soft supersymmetry-breaking terms.

Models obtained from supergravity[\cite{supergravity}]
can provide just such an organizing
principle for the soft supersymmetry-breaking terms specified at some
very high input scale. In minimal supergravity, the spinless particles
in the theory all obtain a common mass $m_0$ at this input scale. In this
scenario, the absence of large flavor-changing neutral currents can be
ascribed to the consequent near-degeneracy of the squarks.  There are also
scalar trilinear couplings among the squarks, sleptons and Higgs scalars
as allowed by $R$-parity and gauge invariance; these are each equal at the
input scale to the corresponding Yukawa coupling times a universal mass
parameter $A$. There is also a supersymmetry-breaking scalar (mass)$^2$
in the Higgs sector. Finally, there are three gaugino masses which also
break supersymmetry. Large CP violation can be avoided with a common
complex phase for all of the soft supersymmetry-breaking parameters.
If one makes further assumptions about the high energy
physics, additional constraints on the parameters of the theory are obtained.
For example, if there is a GUT, then the gauge couplings and gaugino masses
are unified, and there may be relations among the Yukawa couplings and among
the scalar trilinear interactions at the unification scale. However, since the
supersymmetry-breaking mechanism remains mysterious, a precise formulation
of the organizing principle behind the soft terms in the MSSM remains unclear.
$\phantom{\cite{1-13}}$

In any case, with the parameters of the model specified at the input scale
by some candidate organizing principle, one can
run the couplings and masses down to low energies using the renormalization
group equations and make predictions about the sparticle masses and other
low-energy phenomena. Many authors (for example, [\cite{1}-\cite{13}])
have provided numerical and analytical results for
the sparticle spectrum under a variety of assumptions and constraints
on the input parameters. In studies of this type, greater precision as
well as an estimate of errors incurred in the running can be obtained
by employing the two-loop renormalization group equations. The
two-loop $\beta$-functions for the supersymmetric couplings (gauge
couplings[\cite{known}] and superpotential parameters[\cite{yuk}])
have been known for some time.
Recently, the two-loop $\beta$-functions for gaugino mass parameters have
also been found[\cite{us},\cite{yamada}]. In this paper,
we will complete the list of two-loop $\beta$-functions for a general
softly-broken supersymmetric model, by computing the results for
scalar interactions which break supersymmetry softly. We will also
give the results of specializing these calculations to the MSSM.

We consider a general $N=1$ supersymmetric Yang-Mills
model. The chiral superfields $\Phi_i$ contain a complex scalar $\phi_i$
and a two-component fermion $\psi_i$ which transform as a (possibly reducible)
representation $R$ of the gauge group $G$. The superpotential is
$$
W = {1\over 6} Y^{ijk} \Phi_i \Phi_j \Phi_k
+ {1\over 2} \mu^{ij} \Phi_i \Phi_j + L^i \Phi_i \>\>\> .
\eqno(superpotential)
$$
In addition, the Lagrangian contains soft
supersymmetry-breaking terms of the form
$$
{\cal L}_{SB} = -{1\over 6} h^{ijk} \phi_i\phi_j\phi_k
- {1\over 2} \b^{ij} \phi_i \phi_j - {1\over 2}\mij \phi^{*i} \phi_j
- {1\over 2}M \lambda \lambda + {\rm h.c.}
\eqno(soft)
$$
where $M$ is the mass of the gaugino $\lambda$.
The renormalization group equations for the gauge coupling and the
superpotential parameters
$Y^{ijk}$, $\mu^{ij}$ and $L^i$ and the gaugino mass $M$  are known.
In this paper we give the corresponding results for the soft-breaking
parameters $h^{ijk}$, $b^{ij}$, and $(m^2)_i^j$.

For simplicity we first give our results for the special case of a simple
[or $U(1)$] gauge group. We then explain the modifications required if the
gauge group is a direct product in Section III, and discuss
the specialization of these results to the MSSM in Section IV.
We let ${\bf t}^A \equiv ({\bf t})_i^{Aj}$ denote the
representation matrices for the gauge group $G$. Then
$$
\eqalignno{
({\bf t}^A {\bf t}^A)_i^j &\equiv C(R) \delta_i^j &(quadcas)\cr
{\rm Tr}_R ({\bf t}^A {\bf t}^B) &\equiv  S(R) \delta^{AB} &(index) \cr }
$$
define the quadratic Casimir invariant $C(R)$ and the Dynkin index $S(R)$
for the representation $R$.
For the adjoint representation [of dimension denoted by $d(G)$],
$C(G) \delta^{AB} = f^{ACD} f^{BCD}$ with $f^{ABC}$ the
structure constants of the group.

In principle, the two-loop $\beta$-functions for a general renormalizable
theory have already been given in [\cite{MVI},\cite{MVII},\cite{MVIII}].
However, there are two practical issues which must be addressed in order
to apply those results to the case at hand.

The first issue is that only dimensionless couplings appeared in
[\cite{MVI}-\cite{MVIII}]. The two-loop $\beta$-function for a gauge coupling
was given in [\cite{MVI}], for a Yukawa coupling in [\cite{MVII}], and for
a scalar quartic coupling in [\cite{MVIII}]. A general renormalizable theory
may also contain fermion mass terms, scalar (mass)$^2$ terms, and trilinear
scalar couplings. Fortunately, the $\beta$-functions for each of these
dimensionful couplings can be inferred from the results given in [\cite{MVII}]
and [\cite{MVIII}] by taking some of the scalar fields to be non-propagating
``dummy" fields with no gauge interactions. For example,
a fermion mass term has the form
$$
{\cal L}_{\cal M} = -{\cal M}^{ij} \psi_i \psi_j
= -\varphi_{\rm dummy} Y^{ij}_{\rm dummy} \psi_i \psi_j\>\> .
\eqno(dummy)
$$
Now if $\varphi_{\rm dummy}$ is taken to have no other interactions,
then the $\beta$-function for the fermion mass ${\cal M}^{ij}$
has the same form as that of the Yukawa coupling
$Y^{ij}_{\rm dummy}$. Thus the $\beta$-function for any fermion mass
can be inferred directly from the results of [\cite{MVII}]. Similarly, scalar
(mass)$^2$ terms can always
be thought of as scalar quartic interactions involving two
dummy scalars and two normal scalars, while scalar trilinear couplings
can be thought of as scalar quartic interactions with one dummy scalar and
three normal scalars, so the two-loop $\beta$-functions for those cases
can be inferred from the results of [\cite{MVIII}] using a small amount of
careful combinatorics associated with the symmetry factors.

The second issue to be addressed is that the results of
[\cite{MVI}-\cite{MVIII}] were obtained using dimensional regularization
[\cite{dreg}] (DREG). Now,
DREG violates supersymmetry explicitly because it introduces a mismatch
between the numbers of gauge boson and gaugino degrees of freedom.  Therefore
the use of DREG is inappropriate for (even softly broken) supersymmetric
models. Instead, one should use the modified scheme known as dimensional
reduction [\cite{dred}] (DRED) which does not violate supersymmetry.

In DREG, supersymmetry is violated in the finite parts of one-loop graphs, and
in the divergent parts of two-loop graphs. This means that for a given
set of physical quantities (e.g. pole masses and S-matrix elements)
the running couplings computed in DREG with modified minimal
subtraction[\cite{msbar}] ($\msbar$)
will differ from those computed in DRED with
modified minimal subtraction ($\drbar$) by finite one-loop corrections,
and the $\beta$-functions will be different for the two schemes starting
at the two-loop level. We therefore present our $\beta$-functions as they
appear in the $\drbar$ scheme.
This means that the results of [\cite{MVI}-\cite{MVIII}] must be translated
from $\msbar$ to $\drbar$. In [\cite{us}], we provided a
``dictionary" for translating couplings between the two schemes
including all finite one-loop radiative corrections, which is the
order necessary for the present application to two-loop $\beta$-functions.
(The relationship between DRED and DREG for non-supersymmetric theories
has recently been illuminated in [\cite{jjr}].)
We must also take into account the fact that in DRED,
the so-called $\epsilon$-scalars[\cite{dred}] obtain one-loop
mass counterterms due to the supersymmetry-breaking scalar and gaugino masses.

In summary, our method of computation is to use the results of
[\cite{MVI}-\cite{MVIII}] for a general renormalizable theory, using
dummy scalars where necessary for dimensionful couplings, and translating
the results from $\msbar$ to $\drbar$. This process is straightforward
but rather tedious, and we decline to exhibit the details here.
It may also be possible in principle to extract these results from the
dimensional reduction calculations of [\cite{jjo}].

\subhead{II. Two-Loop Running in a General Softly Broken Supersymmetric Model}
\taghead{2.}
 For completeness, we begin by reviewing the known results for gauge couplings,
gaugino masses, and superpotential parameters (including Yukawa couplings).
This will also serve as a useful point of reference for the notation
established in the Introduction.
We will then provide the two-loop beta functions for $h^{ijk}$, $b^{ij}$,
and $\mij$, which constitute the new results of this paper.
We do not assume anything about the relative complex phases of any of the
parameters.

The gauge coupling at two loops is actually scheme-independent. Therefore,
it may be obtained simply by specializing the results of [\cite{MVI}]
to a general supersymmetric model.
Doing so, we obtain the known[\cite{known}] result:
$$
\eqalignno{
& {d\over dt} g \> =  \> {1\over 16 \pi^2} \beta_g^{(1)}
+ {1\over (16 \pi^2)^2 } \beta_g^{(2)}
&(betag) \cr
\beta_g^{(1)} \> = \> & g^3 \left[S(R) - 3 C(G) \right]
&(betag1) \cr
\beta_g^{(2)} \> = \> & g^5
\left\{ - 6[C(G)]^2 + 2 C(G) S(R) + 4 S(R) C(R) \right\}
    - g^3 Y^{ijk} Y_{ijk}C(k)/d(G)
\>\> .
&(betag2) }
$$
Here $Y_{ijk} = ( Y^{ijk}  )^*$ and $S(R)$ is the Dynkin index summed
over all chiral multiplets and $S(R)C(R)$ is the sum of the Dynkin indices
weighted by the quadratic Casimir invariant.

The two-loop $\beta$-function for the gaugino mass parameter may be computed
(as we did in [\cite{us}]) by first using the results of [\cite{MVII}] for
a general theory in $\msbar$ and then translating the results to the
$\drbar$ scheme appropriate for supersymmetry. The result is:
$$
\eqalignno{
& {d\over dt} M = {1\over 16 \pi^2} \beta_M^{(1)}
+ {1\over (16 \pi^2)^2 } \beta_M^{(2)}
&(betam)\cr
\beta_M^{(1)} = & g^2 \left[ 2 S(R) - 6 C(G) \right] M
&(betam1) \cr
\beta_M^{(2)}
= & g^4\left\{ -24[C(G)]^2 + 8 C(G) S(R) + 16  S(R)C(R)\right\} M\cr
 &\hbox{\hskip 110pt} + 2 g^2 \left [h^{ijk}  - M Y^{ijk}\right] Y_{ijk}
C(k)/d(G) \>\> .  &(betam2)
\cr
}
$$
This result was also obtained by Y.~Yamada[\cite{yamada}]
using a different method.
Note that we have adopted a slightly different notation here than we did
in [\cite{us}], since the heights of indices on the last term in \(betam2)
have been reversed. This is the same as exchanging the role of the gaugino
mass $M$ and its conjugate $M^\dagger$. We do this to ensure consistency with
results below while maintaining the fewest possible number of $M^\dagger$s.

The two-loop $\beta$-functions for the superpotential parameters can
be obtained either by superfield techniques[\cite{yuk}], or by applying the
general results of [\cite{MVII}] and [\cite{MVIII}]. In applying the latter
method, one must be careful to convert the $\msbar$ results into
$\drbar$, as we have already mentioned. We obtain the same results
using both methods:
$$
\eqalignno{
{d\over dt} Y^{ijk} &=  Y^{ijp} \left [
{1\over {16 \pi^2}}\gamma_p^{(1)k} +
{1\over {(16 \pi^2})^2}  \gamma_p^{(2)k} \right ]
+ (k \leftrightarrow i) + (k\leftrightarrow j)
&(betay)\cr
{d\over dt} \mu^{ij} &=  \mu^{ip} \left [
{1\over {16 \pi^2}}\gamma_p^{(1)j} +
{1\over {(16 \pi^2})^2}  \gamma_p^{(2)j} \right ]
+ (j \leftrightarrow i)
&(betamu)\cr
{d\over dt} L^{i} &=  L^{p} \left [
{1\over {16 \pi^2}}\gamma_p^{(1)i} +
{1\over {(16 \pi^2})^2}  \gamma_p^{(2)i} \right ]
&(betal)\cr
}
$$
where
$$
\eqalignno{
\gamma_i^{(1)j} &= {1\over 2} Y_{ipq} Y^{jpq} - 2 \delta_i^j g^2 C(i)
&(gammaone)\cr
\gamma_i^{(2)j} &= -{1\over 2} Y_{imn} Y^{npq} Y_{pqr} Y^{mrj}
+ g^2 Y_{ipq} Y^{jpq} [2C(p)- C(i)]
\cr & \qquad \qquad\qquad
+ 2 \delta_i^j g^4 [ C(i) S(R)+ 2 C(i)^2 - 3 C(G) C(i)]
\>\> .
&(gammatwo)\cr
}
$$
In these equations, $C(r)$ always refers to the quadratic Casimir invariant
of the representation carried by the indicated chiral superfield,
while $S(R)$ refers to the total Dynkin index summed over all of the chiral
superfields. The objects $\gamma_i^{(1)j}$ and
$\gamma_i^{(2)j}$ arise completely from wave-function renormalization
in the superfield approach, in accordance with the ``non-renormalization"
theorems of supersymmetry[\cite{nonren}].

Next we consider the two-loop $\beta$-function for the soft
supersymmetry-breaking scalar trilinear coupling $h^{ijk}$.
This is obtained by
specializing eqs.~(4.3) and (4.7) of [\cite{MVIII}] to the supersymmetric
case, with one of the external scalar fields being a dummy field, and
then translating the result from $\msbar$ to $\drbar$ using [\cite{us}].
The result is:
$$
\eqalignno{
{d\over dt} h^{ijk} \> = \> &
{1\over 16\pi^2} \left [\beta^{(1)}_h\right ]^{ijk} +
{1\over (16\pi^2)^2} \left [\beta^{(2)}_h\right ]^{ijk}
&(betahijk)
\cr
\left [\beta^{(1)}_h\right ]^{ijk}  \> =
\> & {1\over 2} h^{ijl} Y_{lmn} Y^{mnk}
+ Y^{ijl} Y_{lmn} h^{mnk} - 2 \left (h^{ijk} - 2 M Y^{ijk}  \right )
g^2  C(k) \cr
& + (k \leftrightarrow i) + (k \leftrightarrow j) &(betahijk1)\cr
& {}\cr
\left [\beta^{(2)}_h\right ]^{ijk}  \> =
\> &
-{1\over 2} h^{ijl} Y_{lmn} Y^{npq} Y_{pqr} Y^{mrk} \cr
& - Y^{ijl} Y_{lmn} Y^{npq} Y_{pqr} h^{mrk}
- Y^{ijl} Y_{lmn} h^{npq} Y_{pqr} Y^{mrk} \cr
& + \left ( h^{ijl} Y_{lpq} Y^{pqk} +  2 Y^{ijl} Y_{lpq} h^{pqk}
- 2 M Y^{ijl} Y_{lpq} Y^{pqk} \right ) g^2\left[ 2 C(p) - C(k) \right ]   \cr
& + \left (2h^{ijk} - 8 M Y^{ijk} \right )
g^4 \left [  C(k)S(R)+ 2 C(k)^2  - 3 C(G)C(k)\right ] \cr
&+ (k \leftrightarrow i) + (k \leftrightarrow j) \>\> . &(betahijk2)\cr }
$$

Next we consider the two-loop $\beta$-function for the scalar $({\rm mass})^2$
$\b^{ij}$. This is again obtained by
specializing eqs.~(4.3) and (4.7) of [\cite{MVIII}] to the supersymmetric
case, but with two of the external scalar fields as dummy fields, and
then translating the result from $\msbar$ to $\drbar$ using the results
of [\cite{us}]. Doing so, we find:
$$
\eqalignno{
{d\over dt}\b^{ij} \> = \> &
{1\over 16\pi^2} \left [\beta^{(1)}_\b \right ]^{ij} +
{1\over (16\pi^2)^2} \left [\beta^{(2)}_\b \right ]^{ij}
&(betabij) \cr
\left [\beta^{(1)}_\b \right ]^{ij} \>  = \>
& {1\over 2} \b^{il} Y_{lmn} Y^{mnj} +{1\over 2}Y^{ijl} Y_{lmn} \b^{mn}
+ \mu^{il} Y_{lmn} h^{mnj}
- 2 \left (\b^{ij} - 2 M \mu^{ij} \right )g^2 C(i)  \cr
& + (i \leftrightarrow j)  &(betabij1)\cr
&{}\cr
\left [\beta^{(2)}_\b \right ]^{ij} \>  = \>
& -{1\over 2} \b^{il} Y_{lmn} Y^{pqn} Y_{pqr} Y^{mrj}
-{1\over 2} Y^{ijl} Y_{lmn} \b^{mr} Y_{pqr} Y^{pqn} \cr
&-{1\over 2} Y^{ijl} Y_{lmn} \mu^{mr} Y_{pqr} h^{pqn}
- \mu^{il} Y_{lmn} h^{npq} Y_{pqr}  Y^{mrj} \cr
& - \mu^{il} Y_{lmn} Y^{npq}  Y_{pqr} h^{mrj}
+ 2 Y^{ijl} Y_{lpq} \left ( \b^{pq} - \mu^{pq} M \right ) g^2 C(p)
\cr
& + \left ( \b^{il} Y_{lpq} Y^{pqj} + 2 \mu^{il} Y_{lpq} h^{pqj}
- 2 \mu^{il} Y_{lpq} Y^{pqj} M \right )
g^2 \left[ 2 C(p) - C(i) \right ]  \cr
& +  \left ( 2\b^{ij} - 8 \mu^{ij} M\right )
g^4 \left [ C(i)S(R)+ 2 C(i)^2- 3 C(G)C(i)   \right ] \cr
&+ (i \leftrightarrow j) \>\> . &(betabij2) \cr }
$$

Finally, we consider the two-loop $\beta$-function for
the scalar $({\rm mass})^2$ $\mij$. This is once again obtained by
specializing eqs.~(4.3) and (4.7) of [\cite{MVIII}] to the supersymmetric
case, with two of the external scalar fields as dummy fields, and
translating from $\msbar$ to $\drbar$ using [\cite{us}].
One must
also be careful here to take into account the mass counterterms for the
$\epsilon$-scalars which arise in DRED.
The result is:
$$
\eqalignno{
{d\over dt} \mij \> = \> &
{1\over 16\pi^2} \left [\beta^{(1)}_{m^2} \right ]_i^j +
{1\over (16\pi^2)^2} \left [\beta^{(2)}_{m^2} \right ]_i^j
&(mijtwo)
\cr
\left [\beta^{(1)}_{m^2} \right ]_i^j \> = \> &
 {1\over 2} Y_{ipq} Y^{pqn} {(m^2)}_n^j
+ {1\over 2} Y^{jpq} Y_{pqn} {(m^2)}_i^n + 2 Y_{ipq} Y^{jpr} {(m^2)}_r^q
\cr
& + h_{ipq} h^{jpq}  - 8 \delta_i^j M M^\dagger g^2 C(i) +
 2 g^2 {\bf t}^{Aj}_i {\rm Tr} [ {\bf t}^A m^2 ] &(mijtwo1) \cr
&{}\cr
\left [\beta^{(2)}_{m^2} \right ]_i^j \> = \> &
 -{1\over 2} {(m^2)}_i^l Y_{lmn} Y^{mrj} Y_{pqr} Y^{pqn}
 -{1\over 2} {(m^2)}^j_l Y^{lmn} Y_{mri} Y^{pqr} Y_{pqn} \cr
& - Y_{ilm} Y^{jnm} {(m^2)}_r^l Y_{npq} Y^{rpq}
 - Y_{ilm} Y^{jnm} {(m^2)}_n^r Y_{rpq} Y^{lpq}  \cr
& - Y_{ilm} Y^{jnr} {(m^2)}_n^l Y_{pqr} Y^{pqm}
- 2 Y_{ilm} Y^{jln}  Y_{npq} Y^{mpr} {(m^2)}_r^q \cr
& - Y_{ilm} Y^{jln} h_{npq} h^{mpq} - h_{ilm} h^{jln} Y_{npq} Y^{mpq} \cr
& - h_{ilm} Y^{jln} Y_{npq} h^{mpq} - Y_{ilm} h^{jln} h_{npq} Y^{mpq} \cr
& + \biggl [{(m^2)}_i^l Y_{lpq} Y^{jpq}
+ Y_{ipq} Y^{lpq} {(m^2)}_l^j + 4 Y_{ipq} Y^{jpl} {(m^2)}_l^q
+  2 h_{ipq} h^{jpq}
\cr
& \>\>\>\>\>\> - 2 h_{ipq} Y^{jpq} M -2 Y_{ipq} h^{jpq} M^\dagger
+ 4Y_{ipq} Y^{jpq} M M^\dagger
\biggr ]
g^2 \left [C(p) + C(q)- C(i) \right ] \cr
& -2 g^2 {\bf t}^{Aj}_i ({\bf t}^A m^2)_r^l Y_{lpq} Y^{rpq}
+ 8 g^4 {\bf t}^{Aj}_i {\rm Tr} [ {\bf t}^A C(r) m^2 ]  \cr
& + \delta_i^j g^4 M M^\dagger \left [
24C(i) S(R) + 48 C(i)^2 - 72 C(G) C(i) \right ]
\cr & + 8 \delta_i^j g^4 C(i) ( {\rm Tr} [S(r) m^2] - C(G) M M^\dagger )
\>\> . &(betamij2) \cr }
$$
Here $h_{ijk} = (h^{ijk})^*$. The traces are over all of the chiral
superfields, and the $C(r)$ are the quadratic Casimir invariants
for the irreducible representations of chiral superfields
in the traces. Note that the terms which explicitly involve ${\bf t}_i^{Aj}$
are zero for non-abelian groups.

This completes the list of two-loop $\beta$-functions for a general
softly-broken  supersymmetric theory. We conclude this section by noting
a  non-trivial consistency check on these results. It has been
shown[\cite{finite}] that an $N=2$ supersymmetric Yang-Mills theory
is finite to all orders in perturbation theory provided that certain
constraints are imposed on the representations of superfields. This finiteness
continues to hold even with soft breaking, provided
that the soft terms obey certain additional
constraints[\cite{softfinite}]. It is an amusing
exercise to check that when the above formulas for a general $N=1$
supersymmetric model are specialized to the $N=2$ case with the
appropriate contraints on the representations and soft couplings,
the two-loop $\beta$-functions do indeed vanish.

\subhead{III. Two-Loop Beta Functions for Direct Product Groups}
\taghead{3.}

As promised, we now point out the modifications which must be made to the
preceding formulas if the gauge group is a product of simple [or
$U(1)$] subgroups $G_a$.

One obtains the $\beta$-function for each gauge coupling $g_a$ by
applying the following rules to \(betag1)-\(betag2):
$$
\eqalignno{
g^3 C(G) & \rightarrow g_a^3 C(G_a)
&(31)\cr
g^3 S(R) & \rightarrow g_a^3 S_a(R)
&(32)\cr
g^5 C(G)^2 & \rightarrow  g_a^5 C(G_a)^2
&(33)\cr
g^5 C(G)S(R) & \rightarrow g_a^5 C(G_a)S_a(R)
&(34)\cr
g^5 S(R)C(R) & \rightarrow \sum_b g_a^3 g_b^2 S_a(R)C_b(R)
&(35)\cr
g^3 C(k)/d(G) & \rightarrow  g_a^3 C_a(k)/d(G_a)
\>\> . &(36)\cr
}
$$
The $\sum_b$ in \(35) is a sum over subgroups.
Similarly, one obtains the $\beta$-function for each gaugino mass parameter
$M_a$ by applying to \(betam1)-\(betam2) the same rules
as above, but with one less power of $g\rightarrow g_a$, and with
$M\rightarrow M_a$ wherever it appears. The exception is that
$$
16g^4 S(R)C(R) M \rightarrow 8 \sum_b g_a^2 g_b^2 S_a(R)C_b(R)
\> (M_a + M_b)
\eqno(nafta)
$$
in \(betam2). Note that in these cases, all terms which do not contain a
quadratic Casimir invariant of a non-adjoint representation are
simply diagonal in each subgroup.

For the $\beta$-functions of the superpotential parameters
and soft supersymmetry-breaking scalar interactions
\(betay)-\(betamij2), one always obtains a sum over subgroups.
In each of these cases, the following set of rules apply:
$$
\eqalignno{
g^2 C(r) & \rightarrow \sum_a g_a^2 C_a(r) &(37)  \cr
g^4 C(r) S(R) & \rightarrow \sum_a g_a^4 C_a(r) S_a(R)&(38)\cr
g^4 C(r) C(G) & \rightarrow \sum_a g_a^4 C_a(r) C(G_a)&(39)    \cr
g^4 C(r)^2 & \rightarrow \sum_a\sum_b  g_a^2 g_b^2 C_a(r) C_b(r)
\>\> . &(310)\cr }
$$
Terms which also involve gaugino masses are modified exactly as above, with
$$
M,M^\dagger \rightarrow M_a, M_a^\dagger
\eqno(311)
$$
except for the term
$$
48 g^4 M M^\dagger C(i)^2 \rightarrow \sum_a \sum_b g_a^2 g_b^2 C_a(i)
C_b(i) \left [ 32 M_a M_a^\dagger + 8 M_a M_b^\dagger + 8 M_b M_a^\dagger
\right ]
\eqno(312)
$$
in eq.~\(betamij2). We also have in \(mijtwo1)-\(betamij2):
$$\eqalignno{
 g^2 {\bf t}^{Aj}_i {\rm Tr} [ {\bf t}^A m^2 ]
& \rightarrow \sum_a g_a^2 ({\bf t}^A_a)^j_i {\rm Tr} [ {\bf t}^A_a  m^2  ]
&(313) \cr
 g^2 {\bf t}^{Aj}_i ({\bf t}^A m^2)_r^l Y_{lpq} Y^{rpq} & \rightarrow
\sum_a g_a^2 ({\bf t}^A_a)^{j}_{i} ({\bf t}^A_a m^2)_r^l Y_{lpq} Y^{rpq}
&(314) \cr
g^4 {\bf t}^{Aj}_i {\rm Tr} [ {\bf t}^A C(r) m^2 ]
& \rightarrow \sum_a \sum_b
g_a^2 g_b^2 ({\bf t}^A_a)^j_i {\rm Tr} [ {\bf t}^A_a C_b(r) m^2 ]
 &(315) \cr
g^4 C(i) {\rm Tr} [ S(r) m^2] & \rightarrow
\sum_a g_a^4 C_a(i) {\rm Tr} [ S_a(r) m^2]
\>\> . &(316) \cr }
$$
Finally, we must note that there is an exceptional case when the gauge group
contains (unlike the MSSM) a direct product of more than one $U(1)$.
One should then choose the basis for the $U(1)$ subgroups so that the matrix
${\rm Tr}[q_a q_b]$ is diagonal, where the trace is over all chiral superfields
and $q_a$ denotes the $U(1)_a$ charge. This is always possible, since
${\rm Tr}[q_a q_b]$ is always a real symmetric matrix.
Then the only non-trivial rule is that the term $g^4 C(i) {\rm Tr}[ S(r) m^2]$
in \(betamij2) becomes a sum
over non-$U(1)$ subgroups as before, plus a contribution
$$
g^4 C(i) {\rm Tr}[ S(r) m^2]
\rightarrow \sum_a \sum_b g_a^2 g_b^2  (q_a)_i (q_b)_i
{\rm Tr} [q_a q_b m^2  ]
\eqno(ohwhocares)
$$
where $\sum_a$ and $\sum_b$ are sums over $U(1)$ subgroups,
and $(q_a)_i$ denotes the $U(1)_a$ charge of the chiral superfield
carrying the index $i$.

\subhead{IV. Two-Loop Running in the Minimal Supersymmetric Standard Model}
\taghead{4.}

In the MSSM, the gauge group is $SU(3)_c\times SU(2)_{\ss L}
\times U(1)_{\ss Y}$, with chiral superfields $Q$ and $L$ for the
$SU(2)_{\ss L}$-doublet quarks and leptons, and $u$, $d$, $e$ for the
$SU(2)_{\ss L}$-singlet quarks and leptons, and two Higgs doublet chiral
superfields $H_u$ and $H_d$. The superpotential is
$$
W =  u \byuk_u Q H_u+  d \byuk_d Q H_d+  e \byuk_e LH_d + \mu H_u H_d
\eqno(mssmsuper)
$$
where $\byuk_u$, $\byuk_d$, $\byuk_e$ are each
$3\times 3$ Yukawa matrices. The
soft supersymmetry-breaking Lagrangian contains scalar couplings
$$
-{\cal L} =  \sssu \bsh_u \sssq \ssshu+ \sssd \bsh_d \sssq  \ssshd
+ \ssse \bsh_e \sssl \ssshd + B \ssshu \ssshd
+ {\rm h.c.}
\eqno(mssmsoft)
$$
where $\bsh_{u},\bsh_{d},\bsh_{e}$
are again $3\times 3$ matrices in family space,
and a hat is used to denote the scalar component of
each chiral superfield. There are also scalar masses of the $(m^2)_i^j$ type:
$$
-{\cal L} = \mhu \ssshu^\dagger \ssshu + \mhd \ssshd^\dagger \ssshd
+ \sssq^\dagger \bmq \sssq + \sssl^\dagger \bml \sssl +
\sssu \bmu \sssu^\dagger + \sssd\bmd \sssd^\dagger
+ \ssse\bme \ssse^\dagger  \>\> . \eqno(sclrmass)
$$
Here again $\bmq$, $\bml$, $\bmu$, $\bmd$, and $\bme$ are $3\times 3$ matrices
in family space.
Finally, the gauginos for the subgroups $SU(3)_c$, $SU(2)_{\ss L}$ and
$U(1)_{\ss Y}$ have masses $M_3$, $M_2$, and $M_1$, respectively.
Again, we do not assume anything about the complex phases of any of the
parameters. For most applications, it will be sufficient to retain only
the Yukawa couplings of the heaviest family, but we prefer to
retain complete generality.

For the sake of completeness and to provide a useful point of reference,
we begin by reviewing the known results for two-loop $\beta$-functions
within the MSSM. For the three gauge couplings,
we have from \(betag)-\(betag2):
$$
{d\over d t} g_a =  {g_a^3\over 16\pi^2} B^{(1)}_a +
{g_a^3\over (16 \pi^2)^2} \left [
\sum_{b=1}^3 B^{(2)}_{ab} g_b^2 -
\sum_{x=u,d,e} C_a^x \, {\rm Tr }  (Y^\dagger_xY_x) \right ]
\>\> .
\eqno(mssmbetag)
$$
Here $B_a^{(1)} = (33/5,1,-3)$ for $U(1)_{\ss Y}$ (in a GUT normalization),
$SU(2)_{\ss L}$, and  $SU(3)_c$ respectively,  and
$$
B^{(2)}_{ab} = \pmatrix{ {199/ 25} & {27/5} & {88/ 5} \cr
                 {9/ 5}    & 25      & 24          \cr
                 {11/ 5}   &  9          & 14          \cr}
\qquad {\rm and} \qquad
C^{u,d,e}_a = \pmatrix{ {26/ 5} & {14/ 5} & {18/ 5} \cr
                          6         & 6           & 2           \cr
                          4         & 4           & 0           \cr }
\>\> .
\eqno(coeffsbc)
$$
The two-loop
renormalization group equations for the three gaugino mass
parameters [\cite{us},\cite{yamada}]
can then be written easily in terms of the same coefficients:
$$
\eqalign{
{d\over dt} M_a = & {2 g_a^2\over 16\pi^2} B^{(1)}_a M_a +
{2 g_a^2\over (16 \pi^2)^2} \biggl [
\sum_{b=1}^3 B^{(2)}_{ab} g_b^2 (M_a + M_b)
\cr  & \qquad\qquad\qquad
+ \sum_{x=u,d,e} C_a^x \left ( {\rm Tr} [Y_x^\dagger h_x] -
M_a {\rm Tr } [Y_x^\dagger Y_x] \right ) \biggr ]
\>\> .\cr }
\eqno(mssmbetam)
$$

    From \(betay)-\(gammatwo),
the two-loop beta functions for the superpotential parameters are:
$$
{d\over dt} \mu \> = \>  {1\over 16 \pi^2} \beta^{(1)}_{\mu}
                + {1\over (16 \pi^2)^2} \beta^{(2)}_{\mu}
\eqno(mssmmubeta)
$$
$$
{d\over dt} \byuk_{u,d,e} \> =\>
{1\over 16 \pi^2} {\bf \beta}^{(1)}_{\byuk_{u,d,e}}
                + {1\over (16 \pi^2)^2} {\bf \beta}^{(2)}_{\byuk_{u,d,e}}
\eqno(mssmyukbeta)
$$
with:
$$
\eqalignno{
{\beta}^{(1)}_\mu \> = \>
\mu \biggl \lbrace
& {\rm Tr} ( 3\byuk_u \byuk_u^\dagger
+3  \byuk_d \byuk_d^\dagger
+  \byuk_e \byuk_e^\dagger)
- 3 g_2^2 - {3\over 5} g_1^2
\biggr \rbrace
&(mssmmu1)\cr
&{}\cr
{\bf \beta}^{(2)}_\mu \> = \>
\mu \biggl\lbrace
&
- 3 {\rm Tr} (3\byuk_u \byuk_u^\dagger \byuk_u \byuk_u^\dagger
       +3\byuk_d \byuk_d^\dagger \byuk_d \byuk_d^\dagger
       +2\byuk_u \byuk_d^\dagger \byuk_d \byuk_u^\dagger
       +\byuk_e \byuk_e^\dagger \byuk_e \byuk_e^\dagger)
\cr &
+ \Bigl [16 g_3^2 + {4\over 5} g_1^2 \Bigr ]{\rm Tr}(\byuk_u \byuk_u^\dagger )
+ \Bigl [16 g_3^2 - {2\over 5} g_1^2 \Bigr] {\rm Tr}(\byuk_d \byuk_d^\dagger )
+ {6\over 5} g_1^2 {\rm Tr}(\byuk_e \byuk_e^\dagger )
\cr &
+ {15\over 2} g_2^4 + {9\over 5} g_1^2 g_2^2 + {207\over 50} g_1^4
\biggr\rbrace
&(mssmmu2)\cr
}
$$

$$
\eqalignno{
{\bf \beta}^{(1)}_{\byuk_u} \> = \>
\byuk_u \biggl\lbrace
& 3 {\rm Tr} (\byuk_u \byuk_u^\dagger) + 3 \byuk_u^\dagger \byuk_u
+ \byuk_d^\dagger \byuk_d
- {16\over 3} g_3^2 - 3 g_2^2 - {13\over 15} g_1^2
\biggr \rbrace &(mssmbetayu1)
\cr
&{}\cr
{\bf \beta}^{(2)}_{\byuk_u} \> = \>
 \byuk_u \biggl\lbrace
&
- 3{\rm Tr} (3\byuk_u \byuk_u^\dagger \byuk_u \byuk_u^\dagger
+  \byuk_u \byuk_d^\dagger \byuk_d \byuk_u^\dagger)
- \byuk_d^\dagger \byuk_d {\rm Tr} (3 \byuk_d \byuk_d^\dagger
+ \byuk_e \byuk_e^\dagger)
\cr  &
- 9 \byuk_u^\dagger \byuk_u {\rm Tr} (\byuk_u \byuk_u^\dagger)
- 4 \byuk_u^\dagger \byuk_u \byuk_u^\dagger \byuk_u
- 2 \byuk_d^\dagger \byuk_d \byuk_d^\dagger \byuk_d
- 2 \byuk_d^\dagger \byuk_d \byuk_u^\dagger \byuk_u
\cr &
+ \Bigl [ 16 g_3^2 + {4\over 5} g_1^2 \Bigr] {\rm Tr}(\byuk_u \byuk_u^\dagger)
+ \Bigl [ 6 g_2^2 + {2\over 5} g_1^2 \Bigr ] \byuk_u^\dagger \byuk_u
+ {2\over 5} g_1^2 \byuk_d^\dagger \byuk_d
\cr &
-{16\over 9} g_3^4 + 8 g_3^2 g_2^2 + {136\over 45} g_3^2 g_1^2
+ {15\over 2} g_2^4 + g_2^2 g_1^2 + {2743\over 450} g_1^4
\biggr\rbrace   &(mssmbetayu2)
\cr
}
$$

$$
\eqalignno{
{\bf \beta}^{(1)}_{\byuk_d} \> = \>
\byuk_d \biggl\lbrace
& {\rm Tr} (3 \byuk_d \byuk_d^\dagger +\byuk_e \byuk_e^\dagger)
+ 3 \byuk_d^\dagger \byuk_d + \byuk_u^\dagger \byuk_u
- {16\over 3} g_3^2 - 3 g_2^2 - {7\over 15} g_1^2
\biggr \rbrace   &(mssmbetayd1)
\cr
& {} \cr
{\bf \beta}^{(2)}_{\byuk_d} \> = \>
\byuk_d \biggl\lbrace
&
- 3{\rm Tr} (3\byuk_d \byuk_d^\dagger \byuk_d \byuk_d^\dagger
     + \byuk_u \byuk_d^\dagger \byuk_d \byuk_u^\dagger
      + \byuk_e \byuk_e^\dagger \byuk_e \byuk_e^\dagger)
\cr  &
- 3 \byuk_u^\dagger \byuk_u {\rm Tr} (\byuk_u \byuk_u^\dagger)
- 3 \byuk_d^\dagger \byuk_d {\rm Tr} (3 \byuk_d \byuk_d^\dagger
+ \byuk_e \byuk_e^\dagger)
- 4 \byuk_d^\dagger \byuk_d \byuk_d^\dagger \byuk_d
\cr &
- 2 \byuk_u^\dagger \byuk_u \byuk_u^\dagger \byuk_u
- 2 \byuk_u^\dagger \byuk_u \byuk_d^\dagger \byuk_d
+ \Bigl [ 16 g_3^2 -
{2\over 5} g_1^2 \Bigr] {\rm Tr}(\byuk_d \byuk_d^\dagger)
\cr &
+ {6\over 5} g_1^2 {\rm Tr}(\byuk_e \byuk_e^\dagger)
+ {4\over 5} g_1^2 \byuk_u^\dagger \byuk_u
+ \Bigl [6 g^2_2 + {4\over 5} g_1^2 \Bigr] \byuk_d^\dagger \byuk_d
\cr &
-{16\over 9} g_3^4 + 8 g_3^2 g_2^2 + {8\over 9} g_3^2 g_1^2
+ {15\over 2} g_2^4 + g_2^2 g_1^2 + {287\over 90} g_1^4
\biggr\rbrace   &(mssmbetayd2)
\cr
}
$$

$$
\eqalignno{
{\bf \beta}^{(1)}_{\byuk_e} \> = \>
\byuk_e \biggl\lbrace
& {\rm Tr} (3 \byuk_d \byuk_d^\dagger
+ \byuk_e \byuk_e^\dagger) + 3 \byuk_e^\dagger \byuk_e
- 3 g_2^2 - {9\over 5} g_1^2
\biggr \rbrace
&(mssmbetaye1)\cr
&{}\cr
{\bf \beta}^{(2)}_{\byuk_e} \> = \>
\byuk_e \biggl\lbrace
&
- 3 {\rm Tr} (3\byuk_d \byuk_d^\dagger \byuk_d \byuk_d^\dagger
     + \byuk_u \byuk_d^\dagger \byuk_d \byuk_u^\dagger
      + \byuk_e \byuk_e^\dagger \byuk_e \byuk_e^\dagger)
\cr  &
- 3\byuk_e^\dagger \byuk_e {\rm Tr}
(3 \byuk_d \byuk_d^\dagger + \byuk_e \byuk_e^\dagger)
- 4 \byuk_e^\dagger \byuk_e \byuk_e^\dagger \byuk_e
+ \Bigl [ 16 g_3^2 - {2\over  5} g_1^2 \Bigr] {\rm Tr}(\byuk_d \byuk_d^\dagger)
\cr &
+ {6\over  5} g_1^2 {\rm Tr}(\byuk_e \byuk_e^\dagger)
+ 6 g_2^2 \byuk_e^\dagger \byuk_e
+ {15\over 2} g_2^4 + {9\over 5}g_2^2 g_1^2 + {27\over 2} g_1^4
\biggr\rbrace &(mssmbetaye2)
\cr
}
$$
Of course, the $\beta$-functions for $\byuk_{u,d,e}$ are $3\times 3$
matrices in family space.

The above results for the MSSM have all appeared before.
Now we apply our results of Sections II and III to arrive at the
two-loop beta functions for the soft-breaking trilinear scalar couplings:
$$
{d\over dt} \bsh_{u,d,e} = {1\over 16 \pi^2} {\bf \beta}^{(1)}_{\bsh_{u,d,e}}
                + {1\over (16 \pi^2)^2} {\bf \beta}^{(2)}_{\bsh_{u,d,e}}
\>\> .
\eqno(betamssmhude)$$
$$
\eqalignno{
{\bf \beta}^{(1)}_{\bsh_u} \> = \>
\bsh_u \biggl\lbrace
& 3 {\rm Tr} (\byuk_u \byuk_u^\dagger) + 5 \byuk_u^\dagger \byuk_u
+ \byuk_d^\dagger \byuk_d
- {16\over 3} g_3^2 - 3 g_2^2 - {13\over 15} g_1^2
\biggr \rbrace
\cr
+ \byuk_u \biggl\lbrace
& 6 {\rm Tr}(\bsh_u \byuk_u^\dagger) + 4 \byuk_u^\dagger \bsh_u
+ 2 \byuk_d^\dagger \bsh_d
+ {32\over 3} g_3^2 M_3 + 6g_2^2 M_2 + {26\over 15} g_1^2 M_1
\biggr \rbrace
&(mssmbetahu1)\cr
&{}\cr
{\bf \beta}^{(2)}_{\bsh_u} \> = \>
 \bsh_u \biggl\lbrace
&
- 3{\rm Tr} (3\byuk_u \byuk_u^\dagger \byuk_u \byuk_u^\dagger
+  \byuk_u \byuk_d^\dagger \byuk_d \byuk_u^\dagger)
- \byuk_d^\dagger \byuk_d {\rm Tr} (3 \byuk_d \byuk_d^\dagger
+ \byuk_e \byuk_e^\dagger)
\cr  &
- 15 \byuk_u^\dagger \byuk_u {\rm Tr} (\byuk_u \byuk_u^\dagger)
- 6 \byuk_u^\dagger \byuk_u \byuk_u^\dagger \byuk_u
- 2 \byuk_d^\dagger \byuk_d \byuk_d^\dagger \byuk_d
- 4 \byuk_d^\dagger \byuk_d \byuk_u^\dagger \byuk_u
\cr &
+ \Bigl [ 16 g_3^2 + {4\over 5} g_1^2 \Bigr] {\rm Tr}(\byuk_u \byuk_u^\dagger)
+ 12 g_2^2 \byuk_u^\dagger \byuk_u
+ {2\over 5} g_1^2 \byuk_d^\dagger \byuk_d
\cr &
-{16\over 9} g_3^4 + 8 g_3^2 g_2^2 + {136\over 45} g_3^2 g_1^2
+ {15\over 2} g_2^4 + g_2^2 g_1^2 + {2743\over 450} g_1^4
\biggr\rbrace
\cr
 + \byuk_u
\biggl\lbrace
& - 6{\rm Tr} (6\bsh_u \byuk_u^\dagger \byuk_u \byuk_u^\dagger
+ \bsh_u \byuk_d^\dagger \byuk_d \byuk_u^\dagger
+ \bsh_d \byuk_u^\dagger \byuk_u \byuk_d^\dagger)
\cr  &
- 18 \byuk_u^\dagger \byuk_u {\rm Tr} (\bsh_u \byuk_u^\dagger)
- \byuk_d^\dagger \byuk_d {\rm Tr}
(6 \bsh_d \byuk_d^\dagger + 2 \bsh_e \byuk_e^\dagger)
- 12 \byuk_u^\dagger \bsh_u {\rm Tr} (\byuk_u \byuk_u^\dagger)
\cr  &
- \byuk_d^\dagger \bsh_d {\rm Tr} (6 \byuk_d \byuk_d^\dagger
+ 2 \byuk_e \byuk_e^\dagger)
- 6 \byuk_u^\dagger \byuk_u \byuk_u^\dagger \bsh_u
- 8 \byuk_u^\dagger \bsh_u \byuk_u^\dagger \byuk_u
\cr &
- 4 \byuk_d^\dagger \byuk_d \byuk_d^\dagger \bsh_d
- 4 \byuk_d^\dagger \bsh_d \byuk_d^\dagger \byuk_d
- 2 \byuk_d^\dagger \byuk_d \byuk_u^\dagger \bsh_u
- 4 \byuk_d^\dagger \bsh_d \byuk_u^\dagger \byuk_u
\cr &
+ \Bigl [ 32 g_3^2 + {8\over  5} g_1^2 \Bigr] {\rm Tr}(\bsh_u \byuk_u^\dagger)
+ \Bigl [6 g_2^2 + {6 \over 5} g_1^2\Bigr] \byuk_u^\dagger \bsh_u
+ {4\over 5} g_1^2 \byuk_d^\dagger \bsh_d
\cr &
- \Bigl [ 32 g_3^2 M_3 + {8\over  5} g_1^2 M_1\Bigr]
{\rm Tr}(\byuk_u \byuk_u^\dagger)
- \Bigl [12 g_2^2 M_2+ {4\over 5} g_1^2 M_1\Bigr] \byuk_u^\dagger \byuk_u
\cr &
- {4\over 5} g_1^2 M_1 \byuk_d^\dagger \byuk_d
+ {64\over 9} g_3^4 M_3 - 16 g_3^2 g_2^2 (M_3 + M_2)
- {272\over 45} g_3^2 g_1^2 (M_3 + M_1)
\cr &
- 30 g_2^4 M_2 - 2 g_2^2 g_1^2 (M_2 + M_1) - {5486\over 225} g_1^4 M_1
\biggr\rbrace
&(mssmbetahu2)\cr
}
$$

$$
\eqalignno{
{\bf \beta}^{(1)}_{\bsh_d} \> = \>
\bsh_d \biggl\lbrace
& {\rm Tr} (3 \byuk_d \byuk_d^\dagger +\byuk_e \byuk_e^\dagger)
+ 5 \byuk_d^\dagger \byuk_d + \byuk_u^\dagger \byuk_u
- {16\over 3} g_3^2 - 3 g_2^2 - {7\over 15} g_1^2
\biggr \rbrace
&(mssmbetahd1) \cr
+ \byuk_d \biggl\lbrace
& {\rm Tr}(6\bsh_d \byuk_d^\dagger + 2 \bsh_e \byuk_e^\dagger)
+ 4 \byuk_d^\dagger \bsh_d + 2 \byuk_u^\dagger \bsh_u
+ {32\over 3} g_3^2 M_3 + 6g_2^2 M_2 + {14\over 15} g_1^2 M_1
\biggr\rbrace
\cr
& {} \cr
{\bf \beta}^{(2)}_{\bsh_d} \> = \>
\bsh_d \biggl\lbrace
&
- 3{\rm Tr} (3\byuk_d \byuk_d^\dagger \byuk_d \byuk_d^\dagger
        + \byuk_u \byuk_d^\dagger \byuk_d \byuk_u^\dagger
          + \byuk_e \byuk_e^\dagger \byuk_e \byuk_e^\dagger)
\cr  &
- 3 \byuk_u^\dagger \byuk_u {\rm Tr} (\byuk_u \byuk_u^\dagger)
- 5 \byuk_d^\dagger \byuk_d {\rm Tr} (3 \byuk_d \byuk_d^\dagger
+ \byuk_e \byuk_e^\dagger)
- 6 \byuk_d^\dagger \byuk_d \byuk_d^\dagger \byuk_d
\cr &
- 2 \byuk_u^\dagger \byuk_u \byuk_u^\dagger \byuk_u
- 4 \byuk_u^\dagger \byuk_u \byuk_d^\dagger \byuk_d
+ \Bigl [ 16 g_3^2 -
{2\over 5} g_1^2 \Bigr] {\rm Tr}(\byuk_d \byuk_d^\dagger)
\cr &
+ {6\over 5} g_1^2 {\rm Tr}(\byuk_e \byuk_e^\dagger)
+ {4\over 5} g_1^2 \byuk_u^\dagger \byuk_u
+ \Bigl [12 g^2_2 + {6\over 5} g_1^2 \Bigr] \byuk_d^\dagger \byuk_d
\cr &
-{16\over 9} g_3^4 + 8 g_3^2 g_2^2 + {8\over 9} g_3^2 g_1^2
+ {15\over 2} g_2^4 + g_2^2 g_1^2 + {287\over 90} g_1^4
\biggr\rbrace
\cr
 + \byuk_d
\biggl\lbrace
& - 6{\rm Tr} (6\bsh_d \byuk_d^\dagger \byuk_d \byuk_d^\dagger
    + \bsh_u \byuk_d^\dagger \byuk_d \byuk_u^\dagger
     + \bsh_d \byuk_u^\dagger \byuk_u \byuk_d^\dagger
      + 2\bsh_e \byuk_e^\dagger \byuk_e \byuk_e^\dagger)
\cr  &
- 6 \byuk_u^\dagger \byuk_u {\rm Tr} (\bsh_u \byuk_u^\dagger)
- 6\byuk_d^\dagger \byuk_d {\rm Tr} (3 \bsh_d \byuk_d^\dagger
+ \bsh_e \byuk_e^\dagger)
\cr  &
- 6 \byuk_u^\dagger \bsh_u {\rm Tr} (\byuk_u \byuk_u^\dagger)
- 4 \byuk_d^\dagger \bsh_d {\rm Tr}
(3 \byuk_d \byuk_d^\dagger + \byuk_e \byuk_e^\dagger)
- 6 \byuk_d^\dagger \byuk_d \byuk_d^\dagger \bsh_d
\cr &
- 8 \byuk_d^\dagger \bsh_d \byuk_d^\dagger \byuk_d
- 4 \byuk_u^\dagger \bsh_u \byuk_u^\dagger \byuk_u
- 4 \byuk_u^\dagger \byuk_u \byuk_u^\dagger \bsh_u
- 4 \byuk_u^\dagger \bsh_u \byuk_d^\dagger \byuk_d
\cr &
- 2 \byuk_u^\dagger \byuk_u \byuk_d^\dagger \bsh_d
+ \Bigl [ 32 g_3^2 - {4\over 5} g_1^2 \Bigr] {\rm Tr}(\bsh_d \byuk_d^\dagger)
+ {12\over 5} g_1^2 {\rm Tr}(\bsh_e \byuk_e^\dagger)
+ {8\over 5} g_1^2 \byuk_u^\dagger \bsh_u
\cr &
+ \Bigl [6 g_2^2 + {6\over 5} g_1^2 \Bigr] \byuk_d^\dagger \bsh_d
- \Bigl [ 32 g_3^2 M_3 - {4\over 5} g_1^2 M_1\Bigr]
{\rm Tr}(\byuk_d \byuk_d^\dagger)
- {12\over 5} g_1^2 M_1 {\rm Tr}(\byuk_e \byuk_e^\dagger)
\cr &
- \Bigl [12 g_2^2 M_2 + {8\over 5} g_1^2 M_1\Bigr]
\byuk_d^\dagger \byuk_d - {8\over 5} g_1^2 M_1 \byuk_u^\dagger \byuk_u
+ {64\over 9} g_3^4 M_3 - 16 g_3^2 g_2^2 (M_3 + M_2)
\cr &
- {16\over 9} g_3^2 g_1^2 (M_3 + M_1)
- {30} g_2^4 M_2 - 2 g_2^2 g_1^2 (M_2 + M_1) - {574\over 45} g_1^4 M_1
\biggr\rbrace
&(mssmbetahd2)\cr
}
$$

$$
\eqalignno{
{\bf \beta}^{(1)}_{\bsh_e} \> = \>
\bsh_e \biggl\lbrace
& {\rm Tr} (3 \byuk_d \byuk_d^\dagger
+ \byuk_e \byuk_e^\dagger) + 5 \byuk_e^\dagger \byuk_e
- 3 g_2^2 - {9\over 5} g_1^2
 \biggr \rbrace
\cr
+ \byuk_e \biggl\lbrace
&{\rm Tr}(6 \bsh_d \byuk_d^\dagger + 2\bsh_e \byuk_e^\dagger)
+ 4 \byuk_e^\dagger \bsh_e + 6g_2^2 M_2 + {18\over 5} g_1^2 M_1
\biggr \rbrace
&(mssmbetahe1)\cr
&{}\cr
{\bf \beta}^{(2)}_{\bsh_e} \> = \>
\bsh_e \biggl\lbrace
&
- 3{\rm Tr} (3\byuk_d \byuk_d^\dagger \byuk_d \byuk_d^\dagger
      + \byuk_u \byuk_d^\dagger \byuk_d \byuk_u^\dagger
       +\byuk_e \byuk_e^\dagger \byuk_e \byuk_e^\dagger)
\cr  &
- 5\byuk_e^\dagger \byuk_e {\rm Tr}
(3 \byuk_d \byuk_d^\dagger + \byuk_e \byuk_e^\dagger)
- 6 \byuk_e^\dagger \byuk_e \byuk_e^\dagger \byuk_e
+ \Bigl [ 16 g_3^2 - {2\over  5} g_1^2 \Bigr] {\rm Tr}(\byuk_d \byuk_d^\dagger)
\cr &
+ {6\over  5} g_1^2 {\rm Tr}(\byuk_e \byuk_e^\dagger)
+ \Bigl [12 g_2^2 - {6\over 5} g_1^2 \Bigr] \byuk_e^\dagger \byuk_e
+ {15\over 2} g_2^4 + {9\over 5}g_2^2 g_1^2 + {27\over 2} g_1^4
\biggr\rbrace
\cr
 + \byuk_e
\biggl\lbrace
& - 6{\rm Tr} (6\bsh_d \byuk_d^\dagger \byuk_d \byuk_d^\dagger
      +\bsh_u \byuk_d^\dagger \byuk_d \byuk_u^\dagger
       + \bsh_d \byuk_u^\dagger \byuk_u \byuk_d^\dagger
        + 2\bsh_e \byuk_e^\dagger \byuk_e \byuk_e^\dagger)
\cr &
- 4\byuk_e^\dagger \bsh_e {\rm Tr} (3 \byuk_d \byuk_d^\dagger
+  \byuk_e \byuk_e^\dagger)
- 6\byuk_e^\dagger \byuk_e {\rm Tr} (3 \bsh_d \byuk_d^\dagger
+ \bsh_e \byuk_e^\dagger)
\cr &
- 6 \byuk_e^\dagger \byuk_e \byuk_e^\dagger \bsh_e
- 8 \byuk_e^\dagger \bsh_e \byuk_e^\dagger \byuk_e
\cr &
+\Bigl [ 32 g_3^2 - {4\over 5} g_1^2 \Bigr] {\rm Tr}(\bsh_d \byuk_d^\dagger)
+ {12\over 5} g_1^2 {\rm Tr}(\bsh_e \byuk_e^\dagger)
+ \Bigl [6 g_2^2 + {6\over 5} g_1^2 \Bigr] \byuk_e^\dagger \bsh_e
\cr &
- \Bigl [ 32 g_3^2 M_3 - {4\over 5} g_1^2 M_1\Bigr]
{\rm Tr}(\byuk_d \byuk_d^\dagger)
- {12\over 5} g_1^2 M_1 {\rm Tr}(\byuk_e \byuk_e^\dagger)
-12 g_2^2 M_2 \byuk_e^\dagger \byuk_e
\cr &
- {30} g_2^4 M_2 -{18\over 5}g_2^2 g_1^2 (M_1 + M_2) - 54 g_1^4 M_1
\biggr\rbrace
&(mssmbetahe2)\cr
}
$$
These are again $3\times 3$ matrices in family space. One should note that
there is no universally agreed-upon convention for the sign
of the gaugino masses in these equations, because of the freedom to
rotate the phases of the gaugino mass terms in the Lagrangian.

The MSSM also contains one scalar (mass)$^2$ of the type $b^{ij}$,
which from \(betabij)-\(betabij2) satisfies the
two-loop renormalization group equation:
$$
{d\over dt} B = {1\over 16 \pi^2} {\beta}^{(1)}_{B}
                + {1\over (16 \pi^2)^2} {\beta}^{(2)}_{B}
\eqno(mssmbetab)
$$
$$
\eqalignno{
{\beta}^{(1)}_B \> = \>
B \biggl \lbrace
& {\rm Tr} (3 \byuk_u \byuk_u^\dagger
+3  \byuk_d \byuk_d^\dagger
+ \byuk_e \byuk_e^\dagger)
- 3 g_2^2 - {3\over 5} g_1^2
\biggr \rbrace
\cr
+ \mu \biggl \lbrace
&  {\rm Tr} ( 6 \bsh_u \byuk_u^\dagger
+ 6 \bsh_d \byuk_d^\dagger
+ 2\bsh_e \byuk_e^\dagger)
+ 6 g_2^2 M_2 +{6\over 5} g_1^2 M_1
\biggr \rbrace
&(mssmbetab1)\cr
&{}\cr
{\beta}^{(2)}_B \> = \>
B \biggl\lbrace
&
- 3 {\rm Tr} (3\byuk_u \byuk_u^\dagger \byuk_u \byuk_u^\dagger
   +  3\byuk_d \byuk_d^\dagger \byuk_d \byuk_d^\dagger
   +  2\byuk_u \byuk_d^\dagger \byuk_d \byuk_u^\dagger
   +   \byuk_e \byuk_e^\dagger \byuk_e \byuk_e^\dagger)
\cr &
+ \Bigl [16 g_3^2 + {4\over 5} g_1^2 \Bigr] {\rm Tr}(\byuk_u \byuk_u^\dagger )
+ \Bigl [16 g_3^2 - {2\over 5} g_1^2 \Bigr] {\rm Tr}(\byuk_d \byuk_d^\dagger )
+ {6\over 5} g_1^2 {\rm Tr}(\byuk_e \byuk_e^\dagger )
\cr &
+ {15\over 2} g_2^4 + {9\over 5} g_1^2 g_2^2 + {207\over 50} g_1^4
\biggr\rbrace
\cr
+ \mu \biggl\lbrace
- 12& {\rm Tr} (3 \bsh_u \byuk_u^\dagger \byuk_u \byuk_u^\dagger
+ 3 \bsh_d \byuk_d^\dagger \byuk_d \byuk_d^\dagger
+  \bsh_u \byuk_d^\dagger \byuk_d \byuk_u^\dagger
+  \bsh_d \byuk_u^\dagger \byuk_u \byuk_d^\dagger
+  \bsh_e \byuk_e^\dagger \byuk_e \byuk_e^\dagger)
\cr &
+ \Bigl [32 g_3^2 + {8\over 5} g_1^2 \Bigr] {\rm Tr}(\bsh_u \byuk_u^\dagger )
+ \Bigl [32 g_3^2 - {4\over 5} g_1^2 \Bigr] {\rm Tr}(\bsh_d \byuk_d^\dagger )
+ {12\over 5} g_1^2 {\rm Tr}(\bsh_e \byuk_e^\dagger )
\cr &
- \Bigl [32 g_3^2 M_3 + {8\over 5} g_1^2 M_1 \Bigr]
{\rm Tr}(\byuk_u \byuk_u^\dagger )
- \Bigl [32 g_3^2 M_3 - {4\over 5} g_1^2 M_1 \Bigr]
{\rm Tr}(\byuk_d \byuk_d^\dagger )
\cr &
- {12\over 5} g_1^2 M_1 {\rm Tr}(\byuk_e \byuk_e^\dagger )
- {30} g_2^4 M_2 - {18\over 5} g_1^2 g_2^2 (M_1 + M_2) - {414\over 25}g_1^4 M_1
\biggr\rbrace \>\> .
&(mssmbetab2)\cr
}
$$

Finally, we turn to the $\beta$-functions for the scalar (mass)$^2$ terms
of the $\mij$ type in the MSSM. It is convenient to define the quantities
$$
\trym \> = \>
\mhu - \mhd + {\rm Tr} [\bmq - \bml - 2 \bmu+ \bmd  + \bme]
\eqno(calsdef)
$$
and
$$
\eqalignno{
\sss \> = \>
&{\rm Tr} \Bigl [
-(3 \mhu + \bmq) \byuk_u^\dagger \byuk_u + 4 \byuk_u^\dagger \bmu \byuk_u
+ (3\mhd - \bmq) \byuk_d^\dagger \byuk_d - 2 \byuk_d^\dagger \bmd \byuk_d
\cr
&\qquad\qquad
+ (\mhd + \bml) \byuk_e^\dagger \byuk_e - 2 \byuk_e^\dagger \bme \byuk_e
\Bigr ]
\cr &+ \left [ {3\over 2} g_2^2 + {3\over 10}g_1^2 \right ]
\left\lbrace \mhu - \mhd - {\rm Tr} (\bml) \right \rbrace
+ \left [ {8\over 3} g_3^2 + {3\over 2} g_2^2 + {1\over 30} g_1^2 \right ]
{\rm Tr} (\bmq )
\cr
&-\left [ {16\over 3} g_3^2 + {16\over 15} g_1^2 \right ]
{\rm Tr} (\bmu )
+\left [ {8\over 3} g_3^2 + {2\over 15} g_1^2 \right ]
{\rm Tr} (\bmd )
+ {6\over 5} g_1^2 {\rm Tr} (\bme)
&(calsprdef)\cr
}
$$
and
$$
\eqalignno{
\sigma_1 \> = \> &{1\over 5} g_1^2
\Bigl \lbrace
3 ( \mhu + \mhd ) + {\rm Tr}
[\bmq + 3 \bml + 8 \bmu + 2 \bmd + 6 \bme ]\Bigr \rbrace
\cr
\sigma_2 \> =\> & g_2^2
\left \lbrace
\mhu + \mhd  + {\rm Tr}
[3 \bmq +  \bml ]\right \rbrace
\cr
\sigma_3 \> = \> & g_3^2
{\rm Tr} [2 \bmq +  \bmu + \bmd ]  \>\> .
\cr
}
$$

Then from the results of sections II and III,
we obtain for each of the $\beta$-functions in the standard form
$$
{d\over dt} m^2 = {1\over 16\pi^2} \beta^{(1)}_{m^2} +
{1\over {(16\pi^2)^2}} \beta^{(2)}_{m^2}
\eqno(genericbetam2)
$$
the following results:
$$
\eqalignno{
\beta^{(1)}_{\mhu}
\> = \>
& 6 {\rm Tr} [(\mhu + \bmq)\byuk_u^\dagger \byuk_u
+ \byuk_u^\dagger \bmu \byuk_u
+ \bsh_u^\dagger \bsh_u ]
\cr
&- 6 g_2^2 |M_2|^2 - {6\over 5} g_1^2 |M_1|^2
+ {3\over 5} g_1^2 \trym
&(betamhu1)\cr {}\cr
\beta^{(2)}_{\mhu}
\> = \>
&-6 {\rm Tr} \biggl [
 6 (\mhu + \bmq) \byuk_u^\dagger \byuk_u \byuk_u^\dagger \byuk_u
+ 6 \byuk_u^\dagger \bmu \byuk_u \byuk_u^\dagger \byuk_u
\cr
&\qquad\>\>+ (\mhu+\mhd+\bmq) \byuk_u^\dagger \byuk_u \byuk_d^\dagger \byuk_d
 + \byuk_u^\dagger \bmu \byuk_u \byuk_d^\dagger \byuk_d
\cr
&\qquad\>\>+ \byuk_u^\dagger  \byuk_u \bmq\byuk_d^\dagger \byuk_d
+  \byuk_u^\dagger  \byuk_u \byuk_d^\dagger \bmd \byuk_d
+6 \bsh^\dagger_u \bsh_u \byuk_u^\dagger \byuk_u
+6 \bsh^\dagger_u \byuk_u \byuk_u^\dagger \bsh_u
\cr
&\qquad\>\>+ \bsh^\dagger_d \bsh_d \byuk_u^\dagger \byuk_u
+ \byuk^\dagger_d \byuk_d \bsh_u^\dagger \bsh_u
+\bsh^\dagger_d \byuk_d \byuk_u^\dagger \bsh_u
+\byuk^\dagger_d \bsh_d \bsh_u^\dagger \byuk_u \biggr ]
\cr &+\Bigl [ 32 g_3^2 + {8\over 5}g_1^2 \Bigr ]
{\rm Tr} [(\mhu + \bmq)  \byuk^\dagger_u \byuk_u
+ \byuk^\dagger_u \bmu \byuk_u
+ \bsh_u^\dagger \bsh_u ]
\cr
& + 32 g_3^2 \Bigl \lbrace 2 |M_3|^2 {\rm Tr}
[\byuk_u^\dagger \byuk_u]
- M_3^* {\rm Tr} [ \byuk_u^\dagger\bsh_u  ]
- M_3 {\rm Tr} [\bsh_u^\dagger \byuk_u  ]
\Bigr \rbrace
\cr
& + {8\over 5} g_1^2 \Bigl \lbrace 2 |M_1|^2 {\rm Tr}
[\byuk_u^\dagger \byuk_u]
- M_1^* {\rm Tr} [ \byuk_u^\dagger \bsh_u ]
- M_1 {\rm Tr} [  \bsh_u^\dagger\byuk_u ]
\Bigr \rbrace
+ {6\over 5} g_1^2  \sss
\cr &
+ 33 g_2^4 |M_2|^2
+ {18\over 5} g_2^2 g_1^2 (|M_2|^2 + |M_1|^2 + {\rm Re}[M_1 M_2^*])
+{621\over 25} g_1^4 |M_1|^2
\cr & + 3 g_2^2 \sigma_2 + {3\over 5} g_1^2 \sigma_1
&(betamhu2)\cr}
$$

$$
\eqalignno{
\beta^{(1)}_{\mhd}
\> = \>
& {\rm Tr} \Bigl [6(\mhd + \bmq) \byuk_d^\dagger \byuk_d
   + 6\byuk_d^\dagger \bmd  \byuk_d
+2(\mhd + \bml)\byuk_e^\dagger \byuk_e
+ 2 \byuk_e^\dagger \bme \byuk_e
\cr
&\qquad+ 6 \bsh_d^\dagger \bsh_d
+ 2 \bsh_e^\dagger \bsh_e \Bigr]
- 6 g_2^2 |M_2|^2 - {6\over 5} g_1^2 |M_1|^2
- {3\over 5} g_1^2 \trym
&(betamhd1)\cr
&{}\cr
\beta^{(2)}_{\mhd}
\> = \>
&-6 {\rm Tr} \biggl [
6(\mhd+\bmq) \byuk_d^\dagger \byuk_d \byuk_d^\dagger \byuk_d
+ 6 \byuk_d^\dagger \bmd \byuk_d \byuk_d^\dagger \byuk_d
\cr
&\qquad\>\>
+ (\mhu+\mhd + \bmq) \byuk_u^\dagger \byuk_u \byuk_d^\dagger \byuk_d
+ \byuk_u^\dagger \bmu \byuk_u \byuk_d^\dagger \byuk_d
\cr
&\qquad\>\> + \byuk_u^\dagger  \byuk_u \bmq\byuk_d^\dagger \byuk_d
+ \byuk_u^\dagger  \byuk_u \byuk_d^\dagger \bmd \byuk_d
+ 2 (\mhd + \bml)\byuk_e^\dagger \byuk_e \byuk_e^\dagger \byuk_e
\cr
&\qquad\>\> + 2 \byuk_e^\dagger \bme \byuk_e \byuk_e^\dagger \byuk_e
+ 6 \bsh^\dagger_d \bsh_d \byuk_d^\dagger \byuk_d
+ 6 \bsh^\dagger_d \byuk_d \byuk_d^\dagger \bsh_d
+ \bsh^\dagger_u \bsh_u \byuk_d^\dagger \byuk_d
\cr
&\qquad\>\> + \byuk^\dagger_u \byuk_u \bsh_d^\dagger \bsh_d
+ \bsh^\dagger_u \byuk_u \byuk_d^\dagger \bsh_d
+\byuk^\dagger_u \bsh_u \bsh_d^\dagger \byuk_d
+ 2 \bsh^\dagger_e \bsh_e \byuk_e^\dagger \byuk_e
+ 2 \bsh^\dagger_e \byuk_e \byuk_e^\dagger \bsh_e \biggr ]
\cr &+\Bigl [ 32 g_3^2 - {4\over 5}g_1^2 \Bigr ]
{\rm Tr} [ (\mhd + \bmq)\byuk^\dagger_d \byuk_d
+ \byuk^\dagger_d \bmd \byuk_d
+ \bsh_d^\dagger \bsh_d ]
\cr
& + 32 g_3^2 \Bigl \lbrace 2 |M_3|^2 {\rm Tr}
[\byuk_d^\dagger \byuk_d]
- M_3^* {\rm Tr} [ \byuk_d^\dagger\bsh_d  ]
- M_3 {\rm Tr} [\bsh_d^\dagger \byuk_d  ]
\Bigr \rbrace
\cr
& - {4\over 5} g_1^2 \Bigl \lbrace 2 |M_1|^2 {\rm Tr}
[\byuk_d^\dagger \byuk_d]
- M_1^* {\rm Tr} [ \byuk_d^\dagger \bsh_d ]
- M_1 {\rm Tr} [  \bsh_d^\dagger\byuk_d ]
\Bigr \rbrace
\cr
&+ {12\over 5} g_1^2 \Bigl \lbrace
{\rm Tr} [ (\mhd + \bml)\byuk_e^\dagger\byuk_e
+  \byuk_e^\dagger \bme \byuk_e
+\bsh_e^\dagger \bsh_e ]
+2 |M_1|^2 {\rm Tr}[\byuk_e^\dagger \byuk_e ]
\cr
&\qquad\qquad\>\>\>
- M_1 {\rm Tr}[\bsh_e^\dagger \byuk_e ]
- M^*_1 {\rm Tr}[\byuk_e^\dagger \bsh_e ]
\Bigr \rbrace
- {6\over 5} g_1^2  \sss
+ 33 g_2^4 |M_2|^2
\cr &
+ {18\over 5} g_2^2 g_1^2 (|M_2|^2 + |M_1|^2 + {\rm Re}[M_1 M_2^*])
+{621\over 25} g_1^4 |M_1|^2
\cr &+ 3 g_2^2 \sigma_2 + {3\over 5} g_1^2 \sigma_1
&(betamhd2) \cr}
$$
The $\beta$-functions for $\bmq,\bml,\bmu,\bmd,\bme$ are of course
$3\times 3$ matrices:
$$
\eqalignno{
\beta^{(1)}_{\bmq}
\> = \>
& (\bmq + 2 \mhu ) \byuk_u^\dagger \byuk_u
+ (\bmq + 2 \mhd ) \byuk_d^\dagger \byuk_d
+ [ \byuk_u^\dagger \byuk_u + \byuk_d^\dagger \byuk_d ] \bmq
+ 2 \byuk_u^\dagger \bmu \byuk_u
\cr &
+ 2 \byuk_d^\dagger \bmd \byuk_d
+ 2 \bsh_u^\dagger \bsh_u + 2 \bsh_d^\dagger \bsh_d
- {32\over 3} g_3^2 |M_3|^2 - 6 g_2^2 |M_2|^2 - {2\over 15} g_1^2 |M_1|^2
+ {1\over 5} g_1^2 \trym
\cr & {} &(betamq1)\cr
&{}\cr
\beta^{(2)}_{\bmq}
\> = \>
& -(2 \bmq + 8 \mhu ) \byuk_u^\dagger \byuk_u\byuk_u^\dagger \byuk_u
-4  \byuk_u^\dagger \bmu \byuk_u\byuk_u^\dagger \byuk_u
-4  \byuk_u^\dagger  \byuk_u \bmq  \byuk_u^\dagger \byuk_u
\cr &
-4  \byuk_u^\dagger  \byuk_u \byuk_u^\dagger \bmu \byuk_u
-2  \byuk_u^\dagger  \byuk_u \byuk_u^\dagger \byuk_u \bmq
-(2 \bmq + 8 \mhd ) \byuk_d^\dagger \byuk_d\byuk_d^\dagger \byuk_d
\cr &
-4  \byuk_d^\dagger \bmd \byuk_d\byuk_d^\dagger \byuk_d
-4  \byuk_d^\dagger  \byuk_d \bmq  \byuk_d^\dagger \byuk_d
-4  \byuk_d^\dagger  \byuk_d \byuk_d^\dagger \bmd \byuk_d
-2  \byuk_d^\dagger  \byuk_d \byuk_d^\dagger \byuk_d \bmq
\cr
&- \Bigl [ (\bmq + 4 \mhu )\byuk_u^\dagger \byuk_u
+ 2 \byuk_u^\dagger \bmu \byuk_u +  \byuk_u^\dagger \byuk_u \bmq
\Bigr ] {\rm Tr}(3 \byuk_u^\dagger \byuk_u )
\cr
&- \Bigl [ (\bmq + 4 \mhd )\byuk_d^\dagger \byuk_d
+ 2 \byuk_d^\dagger \bmd \byuk_d +  \byuk_d^\dagger \byuk_d \bmq
\Bigr ] {\rm Tr}(3 \byuk_d^\dagger \byuk_d + \byuk_e^\dagger \byuk_e )
\cr
& - 6\byuk_u^\dagger \byuk_u {\rm Tr} ( \bmq\byuk_u^\dagger \byuk_u  +
\byuk_u^\dagger \bmu \byuk_u)
\cr
& - \byuk_d^\dagger \byuk_d {\rm Tr} (6 \bmq\byuk_d^\dagger \byuk_d  +
6 \byuk_d^\dagger \bmd \byuk_d + 2 \bml \byuk_e^\dagger \byuk_e +
2 \byuk_e^\dagger \bme \byuk_e )
\cr
&
- 4 \Bigl \lbrace \byuk_u^\dagger \byuk_u \bsh_u^\dagger \bsh_u
+ \bsh_u^\dagger \bsh_u \byuk_u^\dagger \byuk_u
+ \byuk_u^\dagger \bsh_u \bsh_u^\dagger \byuk_u
+ \bsh_u^\dagger \byuk_u \byuk_u^\dagger \bsh_u   \Bigr\rbrace
\cr
& - 4 \Bigl\lbrace \byuk_d^\dagger \byuk_d \bsh_d^\dagger \bsh_d
+ \bsh_d^\dagger \bsh_d \byuk_d^\dagger \byuk_d
+ \byuk_d^\dagger \bsh_d \bsh_d^\dagger \byuk_d
+ \bsh_d^\dagger \byuk_d \byuk_d^\dagger \bsh_d    \Bigr\rbrace
\cr
&
- \bsh_u^\dagger \bsh_u  {\rm Tr}[ 6 \byuk_u^\dagger \byuk_u ]
- \byuk_u^\dagger \byuk_u  {\rm Tr}[ 6 \bsh_u^\dagger \bsh_u ]
- \bsh_u^\dagger \byuk_u  {\rm Tr}[ 6 \byuk_u^\dagger \bsh_u ]
- \byuk_u^\dagger \bsh_u  {\rm Tr}[ 6 \bsh_u^\dagger \byuk_u ]
\cr
&
- \bsh_d^\dagger \bsh_d  {\rm Tr}[ 6 \byuk_d^\dagger \byuk_d +
2 \byuk_e^\dagger \byuk_e ]
- \byuk_d^\dagger \byuk_d  {\rm Tr}[ 6 \bsh_d^\dagger \bsh_d +
2 \bsh_e^\dagger \bsh_e ]
\cr
&
- \bsh_d^\dagger \byuk_d  {\rm Tr}[ 6 \byuk_d^\dagger \bsh_d +
2 \byuk_e^\dagger \bsh_e ]
- \byuk_d^\dagger \bsh_d  {\rm Tr}[ 6 \bsh_d^\dagger \byuk_d +
2 \bsh_e^\dagger \byuk_e ]
\cr
&+ {2\over 5} g_1^2 \biggl \lbrace
(2\bmq + 4 \mhu )\byuk_u^\dagger \byuk_u
+ 4 \byuk_u^\dagger \bmu \byuk_u
+ 2 \byuk_u^\dagger \byuk_u \bmq
+ 4 \bsh_u^\dagger \bsh_u
 - 4 M_1\bsh_u^\dagger \byuk_u
\cr
& \qquad\qquad
- 4 M_1^* \byuk_u^\dagger \bsh_u  + 8 |M_1|^2 \byuk_u^\dagger \byuk_u
+(\bmq + 2 \mhd )\byuk_d^\dagger \byuk_d
+ 2 \byuk_d^\dagger \bmd \byuk_d
\cr
&\qquad\qquad+  \byuk_d^\dagger \byuk_d \bmq
+ 2 \bsh_d^\dagger \bsh_d
- 2 M_1\bsh_d^\dagger \byuk_d
- 2 M_1^* \byuk_d^\dagger \bsh_d  + 4 |M_1|^2 \byuk_d^\dagger \byuk_d
\biggr \rbrace
\cr &
+ {2\over 5} g_1^2 \sss - {128\over 3} g_3^4 |M_3|^2
+ 32 g_3^2 g_2^2 (|M_3|^2 + |M_2|^2 + {\rm Re}[M_2 M_3^*])
\cr &
+ {32\over 45} g_3^2 g_1^2 (|M_3|^2 + |M_1|^2 + {\rm Re}[M_1 M_3^*])
+ 33 g_2^4 |M_2|^2
\cr &
+ {2\over 5} g_2^2 g_1^2 (|M_2|^2 + |M_1|^2 + {\rm Re}[M_1 M_2^*])
+{199\over 75} g_1^4 |M_1|^2
\cr & + {16\over3} g_3^2 \sigma_3 + 3 g_2^2 \sigma_2 + {1\over 15}
g_1^2\sigma_1
&(betamq2)\cr}
$$

$$
\eqalignno{
\beta^{(1)}_{\bml}
\> = \>
& (\bml + 2 \mhd ) \byuk_e^\dagger \byuk_e
+ 2 \byuk_e^\dagger \bme \byuk_e
+ \byuk_e^\dagger \byuk_e  \bml
+ 2 \bsh_e^\dagger \bsh_e
\cr &- 6 g_2^2 |M_2|^2- {6\over 5} g_1^2 |M_1|^2
- {3\over 5} g_1^2 \trym
&(betaml1)\cr
&{}\cr
\beta^{(2)}_{\bml}
\> = \>
& -(2 \bml + 8 \mhd ) \byuk_e^\dagger \byuk_e\byuk_e^\dagger \byuk_e
-4  \byuk_e^\dagger \bme \byuk_e\byuk_e^\dagger \byuk_e
-4  \byuk_e^\dagger  \byuk_e \bml \byuk_e^\dagger \byuk_e
\cr &
-4  \byuk_e^\dagger \byuk_e\byuk_e^\dagger \bme \byuk_e
-2  \byuk_e^\dagger \byuk_e\byuk_e^\dagger  \byuk_e \bml
\cr &
- \Bigl [ (\bml + 4 \mhd )\byuk_e^\dagger \byuk_e
+ 2 \byuk_e^\dagger \bme \byuk_e +  \byuk_e^\dagger \byuk_e \bml
\Bigr ] {\rm Tr} (3 \byuk_d^\dagger \byuk_d + \byuk_e^\dagger \byuk_e )
\cr
& - \byuk_e^\dagger \byuk_e {\rm Tr} [6 \bmq \byuk_d^\dagger \byuk_d  +
6\byuk_d^\dagger \bmd \byuk_d + 2\bml\byuk_e^\dagger \byuk_e +
2\byuk_e^\dagger \bme \byuk_e ]
\cr
&
- 4 \Bigl\lbrace \byuk_e^\dagger \byuk_e \bsh_e^\dagger \bsh_e
+ \bsh_e^\dagger \bsh_e \byuk_e^\dagger \byuk_e
+ \byuk_e^\dagger \bsh_e \bsh_e^\dagger \byuk_e
+ \bsh_e^\dagger \byuk_e \byuk_e^\dagger \bsh_e \Bigr\rbrace
\cr
&
- \bsh_e^\dagger \bsh_e  {\rm Tr}[ 6 \byuk_d^\dagger \byuk_d +
2 \byuk_e^\dagger \byuk_e ]
- \byuk_e^\dagger \byuk_e  {\rm Tr}[ 6 \bsh_d^\dagger \bsh_d +
2 \bsh_e^\dagger \bsh_e ]
\cr
&
- \bsh_e^\dagger \byuk_e  {\rm Tr}[ 6 \byuk_d^\dagger \bsh_d +
2 \byuk_e^\dagger \bsh_e ]
- \byuk_e^\dagger \bsh_e  {\rm Tr}[ 6 \bsh_d^\dagger \byuk_d +
2 \bsh_e^\dagger \byuk_e ]
\cr
&+ {6\over 5} g_1^2 \biggl \lbrace
(\bml + 2 \mhd )\byuk_e^\dagger \byuk_e
+ 2 \byuk_e^\dagger \bme \byuk_e
+ \byuk_e^\dagger \byuk_e \bml
+ 2 \bsh_e^\dagger \bsh_e \cr
& \qquad\>\>\>- 2 M_1  \bsh_e^\dagger \byuk_e
- 2 M_1^* \byuk_e^\dagger \bsh_e  + 4 |M_1|^2 \byuk_e^\dagger \byuk_e
\biggr \rbrace
- {6\over 5} g_1^2 \sss
\cr &+ 33 g_2^4 |M_2|^2
+ {18\over 5} g_2^2 g_1^2 (|M_2|^2 + |M_1|^2 + {\rm Re}[M_1 M_2^*])
+{621\over 25} g_1^4 |M_1|^2
\cr &+ 3 g_2^2 \sigma_2 + {3\over 5} g_1^2 \sigma_1
&(betaml2)\cr }
$$

$$
\eqalignno{
\beta^{(1)}_{\bmu}
\> = \>
& (2\bmu + 4 \mhu ) \byuk_u \byuk_u^\dagger
+ 4\byuk_u \bmq\byuk_u^\dagger
+ 2\byuk_u \byuk_u^\dagger  \bmu
+ 4 \bsh_u \bsh^\dagger_u
\cr
&- {32\over 3} g_3^2 |M_3|^2 - {32\over 15} g_1^2 |M_1|^2
- {4\over 5} g_1^2 \trym
&(betamu1)\cr
&{}\cr
\beta^{(2)}_{\bmu}
\> = \>
& -(2 \bmu + 8 \mhu ) \byuk_u \byuk^\dagger_u\byuk_u \byuk^\dagger_u
-4 \byuk_u \bmq \byuk^\dagger_u\byuk_u \byuk^\dagger_u
-4 \byuk_u  \byuk^\dagger_u \bmu \byuk_u \byuk^\dagger_u
\cr &
-4 \byuk_u  \byuk^\dagger_u  \byuk_u \bmq \byuk^\dagger_u
-2 \byuk_u  \byuk^\dagger_u  \byuk_u \byuk^\dagger_u \bmu
-(2 \bmu + 4 \mhu + 4 \mhd) \byuk_u \byuk^\dagger_d\byuk_d \byuk^\dagger_u
\cr
&-4 \byuk_u \bmq \byuk^\dagger_d\byuk_d \byuk^\dagger_u
-4 \byuk_u  \byuk^\dagger_d \bmd \byuk_d \byuk^\dagger_u
-4 \byuk_u  \byuk^\dagger_d  \byuk_d \bmq \byuk^\dagger_u
-2 \byuk_u  \byuk^\dagger_d  \byuk_d \byuk^\dagger_u \bmu
\cr
& - \Bigl [ (\bmu+ 4 \mhu)
\byuk_u \byuk_u^\dagger + 2 \byuk_u \bmq \byuk_u^\dagger
+ \byuk_u \byuk_u^\dagger \bmu \Bigr ]
{\rm Tr }[6\byuk_u^\dagger \byuk_u  ]
\cr
&- 12 \byuk_u \byuk_u^\dagger
{\rm Tr }[\bmq\byuk_u^\dagger \byuk_u  + \byuk_u^\dagger \bmu \byuk_u ]
\cr
&- 4\left \lbrace \bsh_u \bsh_u^\dagger \byuk_u \byuk_u^\dagger
     +\byuk_u \byuk_u^\dagger \bsh_u \bsh_u^\dagger
     +\bsh_u \byuk_u^\dagger \byuk_u \bsh_u^\dagger
     +\byuk_u \bsh_u^\dagger \bsh_u \byuk_u^\dagger  \right \rbrace
\cr
&- 4\left\lbrace \bsh_u \bsh_d^\dagger \byuk_d \byuk_u^\dagger
     +\byuk_u \byuk_d^\dagger \bsh_d \bsh_u^\dagger
     +\bsh_u \byuk_d^\dagger \byuk_d \bsh_u^\dagger
     +\byuk_u \bsh_d^\dagger \bsh_d \byuk_u^\dagger  \right \rbrace
\cr
& -12 \left \lbrace\bsh_u \bsh^\dagger_u {\rm Tr} [ \byuk_u^\dagger \byuk_u]
+\byuk_u \byuk^\dagger_u {\rm Tr} [ \bsh_u^\dagger \bsh_u]
+ \bsh_u \byuk^\dagger_u {\rm Tr} [ \bsh_u^\dagger \byuk_u]
+ \byuk_u \bsh^\dagger_u {\rm Tr} [ \byuk_u^\dagger \bsh_u]\right\rbrace
\cr
&+ \Bigl [ 6 g_2^2 - {2\over 5} g_1^2 \Bigr ] \Bigl \lbrace (\bmu+ 2 \mhu)
\byuk_u\byuk_u^\dagger + 2 \byuk_u \bmq\byuk_u^\dagger
+ \byuk_u\byuk_u^\dagger \bmu + 2 \bsh_u \bsh_u^\dagger \Bigr \rbrace
\cr
& + 12 g_2^2 \Bigl \lbrace 2 |M_2|^2 \byuk_u \byuk_u^\dagger -
M_2^* \bsh_u \byuk_u^\dagger - M_2 \byuk_u \bsh_u^\dagger \Bigr \rbrace
\cr
&- {4\over 5} g_1^2 \Bigl \lbrace 2 |M_1|^2 \byuk_u \byuk_u^\dagger -
M_1^* \bsh_u \byuk_u^\dagger - M_1 \byuk_u \bsh_u^\dagger \Bigr \rbrace
- {8\over 5} g_1^2 \sss\cr
&
- {128\over 3} g_3^4 |M_3|^2
+ {512\over 45} g_3^2 g_1^2 (|M_3|^2 + |M_1|^2 + {\rm Re}[M_1 M_3^*])
+{3424\over 75} g_1^4 |M_1|^2
\cr &+ {16\over 3} g_3^2 \sigma_3 + {16\over 15} g_1^2 \sigma_1
&(betamu2)\cr
}
$$

$$
\eqalignno{
\beta^{(1)}_{\bmd}
\> = \>
& (2\bmd + 4 \mhd ) \byuk_d \byuk_d^\dagger
+ 4\byuk_d \bmq\byuk_d^\dagger
+ 2\byuk_d \byuk_d^\dagger  \bmd
+ 4 \bsh_d \bsh^\dagger_d
\cr
&- {32\over 3} g_3^2 |M_3|^2 - {8\over 15} g_1^2 |M_1|^2
+ {2\over 5} g_1^2 \trym
&(betamd1)\cr
&{}\cr
\beta^{(2)}_{\bmd}
\> = \>
& -(2 \bmd + 8 \mhd ) \byuk_d \byuk^\dagger_d\byuk_d \byuk^\dagger_d
-4 \byuk_d \bmq \byuk^\dagger_d\byuk_d \byuk^\dagger_d
-4 \byuk_d  \byuk^\dagger_d \bmd \byuk_d \byuk^\dagger_d
\cr &
-4 \byuk_d  \byuk^\dagger_d  \byuk_d \bmq \byuk^\dagger_d
-2 \byuk_d \byuk^\dagger_d  \byuk_d \byuk^\dagger_d \bmd
-(2 \bmd + 4 \mhu + 4 \mhd) \byuk_d \byuk^\dagger_u\byuk_u \byuk^\dagger_d
\cr
&-4 \byuk_d \bmq \byuk^\dagger_u\byuk_u \byuk^\dagger_d
-4 \byuk_d  \byuk^\dagger_u \bmu \byuk_u \byuk^\dagger_d
-4 \byuk_d  \byuk^\dagger_u  \byuk_u \bmq \byuk^\dagger_d
-2 \byuk_d  \byuk^\dagger_u  \byuk_u \byuk^\dagger_d \bmd
\cr
& - \Bigl [ (\bmd+ 4 \mhd)
\byuk_d \byuk_d^\dagger + 2 \byuk_d \bmq \byuk_d^\dagger
+ \byuk_d \byuk_d^\dagger \bmd \Bigr ]
{\rm Tr }(6\byuk_d^\dagger \byuk_d+2\byuk_e^\dagger \byuk_e )
\cr
&- 4 \byuk_d \byuk_d^\dagger
{\rm Tr }(3\bmq\byuk_d^\dagger \byuk_d  + 3 \byuk_d^\dagger \bmd \byuk_d
+ \bml \byuk_e^\dagger \byuk_e  + \byuk_e^\dagger \bme \byuk_e )
\cr
&- 4 \Bigl\lbrace \bsh_d \bsh_d^\dagger \byuk_d \byuk_d^\dagger
     +\byuk_d \byuk_d^\dagger \bsh_d \bsh_d^\dagger
     +\bsh_d \byuk_d^\dagger \byuk_d \bsh_d^\dagger
     +\byuk_d \bsh_d^\dagger \bsh_d \byuk_d^\dagger  \Bigr\rbrace
\cr
&- 4 \Bigl\lbrace \bsh_d \bsh_u^\dagger \byuk_u \byuk_d^\dagger
     +\byuk_d \byuk_u^\dagger \bsh_u \bsh_d^\dagger
     +\bsh_d \byuk_u^\dagger \byuk_u \bsh_d^\dagger
     +\byuk_d \bsh_u^\dagger \bsh_u \byuk_d^\dagger  \Bigr\rbrace
\cr
& - 4\bsh_d \bsh^\dagger_d {\rm Tr} ( 3\byuk_d^\dagger \byuk_d +
\byuk_e^\dagger \byuk_e )
-4   \byuk_d \byuk^\dagger_d {\rm Tr} ( 3\bsh_d^\dagger \bsh_d +
\bsh_e^\dagger \bsh_e )
\cr
&-4  \bsh_d \byuk^\dagger_d {\rm Tr} ( 3 \bsh_d^\dagger \byuk_d +
\bsh_e^\dagger \byuk_e )
-4 \byuk_d \bsh^\dagger_d {\rm Tr} ( 3 \byuk_d^\dagger \bsh_d +
\byuk_e^\dagger \bsh_e )
\cr
&+ \Bigl [ 6 g_2^2 + {2\over 5} g_1^2 \Bigr ] \Bigl \lbrace
(\bmd+ 2 \mhd)
\byuk_d\byuk_d^\dagger + 2 \byuk_d \bmq\byuk_d^\dagger
+ \byuk_d\byuk_d^\dagger \bmd + 2 \bsh_d \bsh_d^\dagger \Bigr\rbrace
\cr
& + 12 g_2^2 \Bigl \lbrace 2 |M_2|^2 \byuk_d \byuk_d^\dagger -
M_2^* \bsh_d \byuk_d^\dagger - M_2 \byuk_d \bsh_d^\dagger \Bigr \rbrace
\cr
&+ {4\over 5} g_1^2 \Bigl \lbrace 2 |M_1|^2 \byuk_d \byuk_d^\dagger -
M_1^* \bsh_d \byuk_d^\dagger - M_1 \byuk_d \bsh_d^\dagger \Bigr \rbrace
+ {4\over 5} g_1^2 \sss
\cr
&- {128\over 3} g_3^4 |M_3|^2
+ {128\over 45} g_3^2 g_1^2 (|M_3|^2 + |M_1|^2 + {\rm Re}[M_1 M_3^*])
+{808\over 75} g_1^4 |M_1|^2
\cr &+ {16\over 3} g_3^2 \sigma_3 + {4\over 15} g_1^2 \sigma_1
&(betamd2)\cr
}
$$

$$
\eqalignno{
\beta^{(1)}_{\bme}
\> = \>
& (2\bme + 4 \mhd ) \byuk_e \byuk_e^\dagger
+ 4\byuk_e \bml\byuk_e^\dagger
+ 2\byuk_e \byuk_e^\dagger  \bme
+ 4 \bsh_e \bsh^\dagger_e
\cr
&- {24\over 5} g_1^2 |M_1|^2
+ {6\over 5} g_1^2 \trym
&(betame1)\cr
&{}\cr
\beta^{(2)}_{\bme}
\> = \>
& -(2 \bme + 8 \mhd ) \byuk_e \byuk^\dagger_e\byuk_e \byuk^\dagger_e
-4 \byuk_e \bml \byuk^\dagger_e\byuk_e \byuk^\dagger_e
-4 \byuk_e  \byuk^\dagger_e \bme \byuk_e \byuk^\dagger_e
\cr &
-4 \byuk_e  \byuk^\dagger_e  \byuk_e \bml\byuk^\dagger_e
-2 \byuk_e \byuk^\dagger_e  \byuk_e \byuk^\dagger_e \bme
\cr
& - \Bigl [ (\bme+ 4 \mhd)
\byuk_e \byuk_e^\dagger + 2 \byuk_e \bml \byuk_e^\dagger
+ \byuk_e \byuk_e^\dagger \bme \Bigr ]
{\rm Tr }[6\byuk_d^\dagger \byuk_d+2\byuk_e^\dagger \byuk_e ]
\cr
&- 4 \byuk_e \byuk_e^\dagger
{\rm Tr }[3\bmq\byuk_d^\dagger \byuk_d  + 3 \byuk_d^\dagger \bmd \byuk_d
+ \bml \byuk_e^\dagger \byuk_e + \byuk_e^\dagger \bme \byuk_e ]
\cr
&- 4\Bigl\lbrace \bsh_e \bsh_e^\dagger \byuk_e \byuk_e^\dagger
     +\byuk_e \byuk_e^\dagger \bsh_e \bsh_e^\dagger
     +\bsh_e \byuk_e^\dagger \byuk_e \bsh_e^\dagger
     +\byuk_e \bsh_e^\dagger \bsh_e \byuk_e^\dagger  \Bigr\rbrace
\cr
& - 4\bsh_e\bsh^\dagger_e {\rm Tr} [ 3\byuk_d^\dagger \byuk_d +
\byuk_e^\dagger \byuk_e]
-4   \byuk_e \byuk^\dagger_e {\rm Tr} [ 3\bsh_d^\dagger \bsh_d +
\bsh_e^\dagger \bsh_e]
\cr
&-4  \bsh_e \byuk^\dagger_e {\rm Tr} [3 \bsh_d^\dagger \byuk_d +
\bsh_e^\dagger \byuk_e]
-4 \byuk_e \bsh^\dagger_e {\rm Tr} [3 \byuk_d^\dagger \bsh_d +
\byuk_e^\dagger \bsh_e]
\cr
&+ \Bigl [ 6 g_2^2 - {6\over 5} g_1^2 \Bigr ] \Bigl \lbrace
(\bme+ 2 \mhd)
\byuk_e\byuk_e^\dagger + 2 \byuk_e \bml\byuk_e^\dagger
+ \byuk_e\byuk_e^\dagger \bme + 2 \bsh_e \bsh_e^\dagger \Bigr\rbrace
\cr
& + 12 g_2^2 \Bigl \lbrace 2 |M_2|^2 \byuk_e \byuk_e^\dagger -
M_2^* \bsh_e \byuk_e^\dagger - M_2 \byuk_e \bsh_e^\dagger \Bigr \rbrace
\cr
& -{12\over 5} g_1^2 \Bigl \lbrace 2 |M_1|^2 \byuk_e \byuk_e^\dagger -
M_1^* \bsh_e \byuk_e^\dagger - M_1 \byuk_e \bsh_e^\dagger \Bigr \rbrace
\cr
&+ {12\over 5} g_1^2  \sss
+ {2808\over 25} g_1^4 |M_1|^2
+ {12\over 5} g_1^2 \sigma_1
\>\> . &(betame2) \cr
}
$$

\subhead{V. Conclusion}
\taghead{5.}
In this paper, we have presented the two-loop renormalization group
equations for all couplings in a general softly-broken supersymmetric model,
and in particular for the MSSM. If the sparticles predicted by the MSSM
are found and their spectrum is determined with some accuracy (for example
at an $e^+ e^-$ collider[\cite{jlc}]), these results
may be useful in discriminating between various candidate organizing principles
for the soft supersymmetry-breaking terms at some very high input scale.
One can run the parameters down to low energies to predict
the masses of the sparticle spectrum and other features of low energy
phenomenology in terms of what may turn out to be only a few input parameters.
The masses of the sparticles depend primarily on just those soft
supersymmetry-breaking couplings whose $\beta$-functions have been given
to two loops here. We find that the two-loop $\beta$-functions generally
make a difference of
several percent (compared to the one-loop predictions) for the sparticle
masses, although it is quite difficult to make meaningful estimates of the size
of the two-loop corrections without committing to a specific model. At the
same level of accuracy, one must also be careful to take into account threshold
effects as well as the distinction between running masses and pole masses.
In extensions of the MSSM which have a large non-minimal particle content
above the electroweak scale, the two-loop corrections and threshold effects
are potentially much larger.

{\it Note added.} Since the original preprint version of this paper
appeared, the $\beta$ functions of Sec.~II have been calculated by
Yamada [\cite{yamada2}] using superfield techniques, and by Jack and Jones
[\cite{jj}] working directly in DRED. Our results agreed with theirs
except for discrepancies in the two-loop $\beta$ function (2.20) for
scalar masses of the $(m^2)_i^j$ type, namely the coefficient
of the last term proportional to
$
\delta_i^j g^4 C(i) ( {\rm Tr} [S(r) m^2] - C(G) M M^\dagger )
$
and a possible dependence on the unphysical mass of the $\epsilon$ scalar.
As emphasized in [\cite{jj}], one should properly allow the $\epsilon$
scalars to have masses and mass counterterms in the component field
approach to DRED, and the result given there is correct.
However, one can treat the subtractions in $\drbar$ in such a way
that the two-loop $\beta$ functions do not depend on the unphysical
$\epsilon$ scalar mass, and so that the expressions relating the
pole masses to the running masses of the scalars also do not depend on the
$\epsilon$ scalar masses. (This is related to the prescription used
in [\cite{jj}] by a simple coupling constant redefinition  for the scalar
masses, of the type mentioned there.) The corrected results we have given
here correspond to this prescription, and we are now in agreement with
the authors of [\cite{yamada2}] and [\cite{jj}],
to whom we are grateful for consultations.
The subtleties involved will be reported on elsewhere.
The results in the special case of the MSSM have also been corrected
accordingly.

\subhead{Acknowledgments}
S.P.M. is indebted to Diego Casta\~no for many helpful discussions. M.T.V.
would like to thank the theory group at the University of Southampton
for their hospitality while part of this work was carried out.
We are both grateful to Ian Jack, Tim Jones and Youichi Yamada for
their help in correcting Eq.~(2.20), and we are
further indebted to Jack and Jones for pointing out the correct ordering
of matrices in some of the terms in the
$\beta$ functions of Sec.~IV. This work was supported in part by the
National Science Foundation grants PHY-90-01439 and PHY-93-06906 and
U.~S.~Department of Energy grant DE-FG02-85ER40233.

\references
\doublespace
\hyphenation{Nil-les N-i-e-u-w-e-n-h-u-i-z-e-n N-a-n-o-p-o-u-l-o-s}
\hyphenation{S-u-p-e-r-s-y-m-m-e-t-r-i-c}

\refis{dreg} G.~'t Hooft and M.~Veltman, \np B44, 189, 1972.

\refis{softly} L.~Girardello and M.~T.~Grisaru, \np B194, 65, 1982.

\refis{nonren} J.~Wess and B.~Zumino, \pl 49B, 52, 1974; J.~Iliopoulos
and B.~Zumino, \np B76, 310, 1974; S.~Ferrara, J.~Iliopoulos and B.~Zumino,
\np B77, 413, 1974; B.~Zumino, \np B89, 535, 1975; S.~Ferrara and O.~Piguet,
\np B93, 261, 1975; M.~Grisaru, W.~Siegel, and M.~Rocek, \np B159, 429, 1979.

\refis{unification} P.~Langacker, in Proceedings of the
PASCOS90 Symposium, Eds.~P.~Nath
and S.~Reucroft, (World Scientific, Singapore 1990)
J.~Ellis, S.~Kelley, and D.~Nanopoulos, \pl 260B, 131, 1991;
U.~Amaldi, W.~de Boer, and H.~Furstenau, \pl 260B, 447, 1991;
P.~Langacker and M.~Luo, \pr D44, 817, 1991.

\refis{MVI} M.~E.~Machacek and M.~T.~Vaughn, \np B222, 83, 1983.

\refis{MVII} M.~E.~Machacek and M.~T.~Vaughn, \np B236, 221, 1984.

\refis{MVIII} M.~E.~Machacek and M.~T.~Vaughn, \np B249, 70, 1984.

\refis{finite} P.~S.~Howe, K.~S.~Stelle, and P.~West, \pl 124B, 55, 1983.

\refis{softfinite} A.~Parkes and P.~West, \pl 127B, 353, 1983;
J.-M.~Fr\`ere, L.~Mezincescu, and Y.-P.~Yao,
\pr D29, 1196, 1984;\pr D30, 2238, 1984.

\refis{yuk}  P.~West, \pl 137B, 371, 1984;
D.~R.~T.~Jones and L.~Mezincescu, \pl 138B, 293, 1984.

\refis{dred} W.~Siegel, \pl 84B, 193, 1979; D.~M.~Capper, D.~R.~T.~Jones
and P.~van~Nieuwenhuizen, \np B167, 479, 1980.

\refis{msbar} W.~A.~Bardeen, A.~J.~Buras, D.~W.~Duke and T.~W.~Muta,
\pr D18, 3998, 1978.

\refis{us} S.~P.~Martin and M.~T.~Vaughn,
\pl 318B, 331, 1993.

\refis{reviews}
For reviews, see H.~P.~Nilles,  \prpts 110, 1, 1984 or
H.~E.~Haber and G.~L.~Kane, \prpts 117, 75, 1985.

\refis{supergravity}
A.~Chamseddine, R.~Arnowitt and P.~Nath, \prl 49, 970, 1982;
H.~P.~Nilles, \pl 115B, 193, 1982;
L.~E.~Ib\'a\~nez, \journal Phys.~Lett., 118B, 73, 1982;
R.~Barbieri, S.~Ferrara, and C.~Savoy, \pl 119B, 343, 1982;
L. Hall, J. Lykken and S. Weinberg, \pr D27, 2359, 1983;
P. Nath, R. Arnowitt and A. H. Chamseddine, \np B227, 121, 1983.

\refis{yamada} Y.~Yamada, \prl 72, 25, 1994.

\refis{yamada2} Y.~Yamada, ``Two-loop renormalization group equations
for soft SUSY breaking scalar interactions: supergraph method",
KEK-TH-383, KEK Preprint 93-182, UT-665, January 1994.

\refis{known} D.~R.~T.~Jones, \np B87, 127, 1975;
D.~R.~T.~Jones and L.~Mezincescu, \pl 136B, 242, 1984.

\refis{1} A.~Lahanas and D.~Nanopoulos, \prpts 145, 1, 1987.

\refis{2} G.~G.~Ross and R.~G.~Roberts,\np B377, 571, 1992.

\refis{3} L.~E.~Ib\'a\~nez and  G.~G.~Ross,
``Electroweak Breaking in Supersymmetric Models'',
CERN-TH.6412/92, in {\it Perspectives in Higgs Physics},
edited by G.~Kane (World Scientific, Singapore, 1993).

\refis{4} R. Arnowitt and P. Nath, SSCL-Preprint-229 (1993);
\prl 69, 725, 1992;
\pr D46, 3981, 1992; P.~Nath and R.~Arnowitt,
\pl B289, 368, 1992; ibid, {\bf B287} (1992) 89; \prl 70, 3696, 1993.

\refis{5} S.~Kelley, J.~Lopez, D.~Nanopoulos, H.~Pois, and K.~Yuan,
\pl B273, 423, 1991; \np B398, 3, 1993; J.~Lopez, D.~Nanopoulos, and
A.~Zichichi, \pl, B291, 255, 1992; J.~Lopez, D.~Nanopoulos, and H.Pois,
\pr D47, 2468, 1993.

\refis{6} M.~Drees and M.~Nojiri, \pr D47, 376, 1993.

\refis{7} T.~Elliott, S.~F.~King, and P.~L.~White,
\pl B305, 71, 1993; \pl B314, 56, 1993; ``Radiative Corrections to
Higgs Boson Masses in the Next to Minimal Supersymmetric Standard Model",
SHEP-92/93-21, (hep-ph 9308309).

\refis{8} R.~G.~Roberts and L.~Roszkowski, \pl B309, 329, 1993.

\refis{9} S.~P.~Martin and P.~Ramond,
\pr D48, 5365, 1993.

\refis{10} P.~Ramond, ``Renormalization Group Study of the Minimal
Supersymmetric Standard Model: No Scale Models'', invited talk at workshop
``Recent Advances in the Superworld'', Houston TX, April 1993;
D.~J.~Casta\~no, E.~J.~Piard, and P.~Ramond,
``Renormalization group study of the standard model and its extensions:
II. The minimal supersymmetric standard model." University of Florida
preprint UFIFT-HEP-93-18, (hep-ph 9308335), August 1993.

\refis{11} W.~de Boer, R.~Ehret, and D.~I.~Kazakov, ``Constraints on
SUSY Masses in Supersymmetric Grand Unified Theories", IEKP-KA-93-13,
(hep-ph 9308238), August 1993.

\refis{12} V.~Barger, M.~S.~Berger, and P.~Ohmann, \prd 47, 1093, 1993;
``The Supersymmetric Particle Spectrum", Wisconsin preprint MAD-PH-801,
(hep-ph 9311269), November 1993.

\refis{13} G.~L.~Kane, C.~Kolda, L.~Roszkowski, and J.~D.~Wells,
``Study of Constrained Minimal Supersymmetry", Michigan preprint UM-TH-93-24.

\refis{jlc} See for instance, ``JLC-I", KEK report 92-16, the JLC group,
December 1992, and references contained therein.

\refis{jjr} I.~Jack, D.~R.~T.~Jones and K.~L.~Roberts, \journal
Z. Phys., C62, 161, 1994;
``Equivalence of Dimensional Reduction and Dimensional Regularisation",
Liverpool preprint LTH-329 1994.

\refis{jjo} I.~Jack, \pl 147B, 405, 1984 and I.~Jack and H.~Osborn,
\np B249, 472, 1985.

\refis{jj} I.~Jack and D.~R.~T.~Jones,
``Soft supersymmetry breaking and finiteness" Liverpool preprint LTH-334,
May 1994.

\endreferences\endit\end